\documentclass[a4paper,12pt]{article}
\usepackage{latexsym}
\usepackage{amsmath}
\usepackage{amssymb}
\usepackage{amscd}

\newlength{\minitwocolumn}
\makeatletter
\@addtoreset{equation}{section}
\makeatother

\title{\bf 
The affine $A_{n-1}^{(1)}$ Toda fields\\
with boundary reflection
}

\author{Takeo Kojima}
\date{\it
Department of Mathematics,
College of Science and Technology,\\
Nihon University, Chiyoda-ku, Tokyo
101-0062, Japan\\~\\
E-mail: kojima@math.cst.nihon-u.ac.jp\\~\\
{\rm \today}
}

\begin{document}
\maketitle

\begin{abstract}
We study the affine $A_{n-1}^{(1)}$ 
Toda fields
with boundary reflection.
Our approach is based on the free field 
approach.
We construct
free field realizations of the boundary state and its
dual.
For an application of these realizations,
we present
integral representations for the form factors
of the local operators.
In a limiting case
$\rho \to \infty$,
our integral representations
reproduce those of form factors for
the $SU(n)$
invariant massive Thirring model
with boundary reflection \cite{K1}.
\end{abstract}

\newpage

\section{Introduction}
For without-boundary field theory,
integrability is ensured by
the factorized scattering theory
\cite{ZZ}.
The factorized scattering theory for boundary
field theory
was developed by
\cite{C, GZ, HSY}.
More precisely,
I.Cherednik \cite{C} proposed the boundary Yang-Baxter
equation,
and
S.Ghoshal, A.Zamolodchikov
\cite{GZ} proposed the boundary crossing symmetry.
B.Hou, K.Shi and W.Yang \cite{HSY}
proposed higher rank generalization of
the boundary crossing symmetry.
E.Sklyanin \cite{S} began
Bethe Ansatz treatment and introduced
the commuting transfer matrix for boundary model.
M.Jimbo et al. \cite{JKKKM} developed these ideas for
boundary model,
\cite{C,GZ,S} and
established the free field approach \cite{JKKKM,JM} for
boundary integrable model.
They diagonalized the Hamiltonian
of the massive $XXZ$ model with a boundary,
and derived integral representations for
the spin correlation functions.
The $U_q(\widehat{sl_n})$ generalization
of reference \cite{JKKKM} was achieved
in \cite{FK}.  

In this paper we study the affine $A_{n-1}^{(1)}$
Toda fields with boundary reflection,
by means of the free field approach.
For $A_1^{(1)}$ symmetry case,
Hou et al. \cite{HSWY}
studied the affine $A_1^{(1)}$ Toda
fields (SG model) with boundary reflection,
in terms of the free field approach,
and constructed the boundary state.
They derived integral representations
for form factors of local fields.
In reference \cite{HSWY, FKQ},
constructions of the boundary
state started
with Lukyanov's ultra-violet cut-off realization
of Zamolodchikov-Faddeev operators
\cite{L}, and the form factors were derived after removing
the cut-off parameter at the final stage.
In this paper we prefer to work directly
with operators with cut-off parameter removed,
as the same manner as reference \cite{K2},
in which
the massless $XXZ$ chain with a boundary
was studied.
Using free field realizations
of the Zamolodchikov-Faddeev operators
\cite{MT} and the local operators \cite{KY},
we construct the free field realizations of the boundary
state and its dual, and 
give integral representations for form factors of local
fields of the affine $A_{n-1}^{(1)}$ Toda fields
with boundary reflection.
In this paper we consider the case
where the deformation parameter of quantum
gruop $U_q(\widehat{sl_n})$ is
\begin{eqnarray}
q=e^{-\frac{2\pi^2i}{\rho n}},~\rho>0.
\end{eqnarray}
In a limiting case $\rho \to \infty$,
our integral representations reproduce those of form factors
for the $SU(n)$ invariant massive Thirring model
with boundary reflection \cite{K1}.
For $n=2$ case
(SG model),
it is possible to generalize the boundary condition
of this paper (Appendix C).

Now a few words about the organization of this paper.
In section 2 we set up problem.
In section 3 we present the free field realizations
of Zamolodchikov-Faddeev operators \cite{MT},
and local operators \cite{KY}.
In section 4 we construct the free field realizations
of the boundary state and its dual
(\ref{boundary1}), (\ref{boundary2}).
In section 5 
we derive the boundary qKZ equations
(\ref{BQKZ1}), (\ref{BQKZ2}),
which govern form factors, and
present integral representations
for form factors of local fields
(\ref{1point}), (\ref{special}),
(\ref{main}).
In Appendix A we summarize the multiple gamma functions.
In Appendix B we summarize the contraction relations
of the basic operators.
In Appendix C we summarize a generalization of the boundary
condition for the affine $A_1^{(1)}$ Toda fields
(SG model).

\section{Model}
The purpose of this section is to set up the problem,
and present briefly necessary tools concering the completely integral models of quantum field theory with massive
spectra.
The affine $A_{n-1}^{(1)}$ Toda fields with boundary 
reflection is described by the bulk $S$-matrix and 
the boundary $K$-matrix
\cite{C}, where the amplitudes of the bulk
$S$-matrix and the boundary $K$-matrix are
related by the boundary crossing symmetry \cite{GZ}.
Let $V$ be $n$-dimensional vector space
$V=\oplus_{j=0}^{n-1}{\mathbb{C}}v_j$.
The bulk $S$-matrix $S(\beta) \in {\rm End}(V
\otimes V)$ of the present model is given by
\begin{eqnarray}
S(\beta)v_{j_1}\otimes v_{j_2}
=\sum_{k_1,k_2=0}^{n-1}
v_{k_1}\otimes v_{k_2}
S(\beta)_{k_1,k_2}^{j_1,j_2},\label{def:S}
\end{eqnarray}
where nonzero entries are given by
\begin{eqnarray}
S(\beta)_{j j}^{j j}&=&s(\beta),\\
S(\beta)_{j k}^{j k}&=&
s(\beta)\times
\frac{
\displaystyle
-{\rm sh}\left(\frac{\pi}{\rho}\beta\right)}
{\displaystyle
{\rm sh}\left(\frac{\pi}{\rho}(\beta-\frac{2\pi i}{n})\right)
},~~(j \neq k),
\\
S(\beta)_{j k}^{k j}
&=&s(\beta)\times
\left\{
\begin{array}{cc}
\frac{\displaystyle
-e^{\frac{\pi}{\rho}\beta}
{\rm sh}\frac{2\pi^2 i}{\rho n}}
{\displaystyle
{\rm sh}\left(\frac{\pi}{\rho}(\beta-\frac{2\pi i}{n})\right)
},& (j>k),\\
\frac{\displaystyle
-e^{-\frac{\pi}{\rho}\beta}
{\rm sh}\frac{2\pi^2 i}{\rho n}}
{\displaystyle
{\rm sh}\left(\frac{\pi}{\rho}(\beta-\frac{2\pi i}{n})\right)
},& (j<k),
\end{array}
\right.
\end{eqnarray}
Here we have set 
\begin{eqnarray}
s(\beta)=\frac{S_2(-i\beta|\rho, 2\pi) 
S_2(i\beta+\frac{2(n-1)\pi}{n}|\rho,2\pi)}{
S_2(i\beta|\rho, 2\pi) 
S_2(-i\beta+\frac{2(n-1)\pi}{n}|\rho,2\pi)
}.
\end{eqnarray}
The double trigonometric function
$S_2(x|\omega_1 \omega_2)$ is summarized in Appendix A.
The boundary $K$-matrix $K(\beta) \in {\rm End}(V)$
of the present model is given by
\begin{eqnarray}
K(\beta)v_j=\sum_{k=0}^{n-1}
v_k K(\beta)_k^j,
\end{eqnarray}
where
\begin{eqnarray}
K(\beta)_j^k=\varphi(\beta)^{-1}\times \delta_{j,k},~~~
\varphi(\beta)=\frac{S_2(-2i\beta|\rho,4\pi)
S_2(2i\beta+\frac{2(n-1)\pi}{n}|\rho,4\pi)}{
S_2(2i\beta|\rho,4\pi)
S_2(-2i\beta+\frac{2(n-1)\pi}{n}|\rho,4\pi)
}.\label{def:K}
\end{eqnarray}
The boundary $K$-matrix $K(\beta)$ satisfies 
the boundary Yang-Baxter equation \cite{C},
\begin{eqnarray}
K_2(\beta_2)S_{21}(\beta_1+\beta_2)K_1(\beta_1)
S_{12}(\beta_1-\beta_2)=
S_{21}(\beta_1-\beta_2)K_1(\beta_1)
S_{12}(\beta_1+\beta_2)K_2(\beta_2),
\end{eqnarray}
and the unitary condition,
\begin{eqnarray}
K(\beta)K(-\beta)=id.
\end{eqnarray} 
For general $n\geq 2$ case,
the $K$-matrix $K(\beta)$ satisfies
the boundary crossing symmetry proposed by
\cite{HSY}, in a limiting case $\rho \to \infty$.
Precisely, 
set the matrix $\widetilde{S}(\beta)
\in {\rm End}(V \otimes V)$ and $\widetilde{K}(\beta)
\in {\rm End}(V)$ by
\begin{eqnarray}
\widetilde{S}(\beta)v_{j_1}\otimes v_{j_2}=
\lim_{\rho \to \infty}
\sum_{k_1,k_2=0}^{n-1}
v_{k_1}\otimes v_{k_2}S(\beta)_{k_1,k_2}^{j_1,j_2}
(-1)^{\delta_{j_1,j_2}+\delta_{j_1,k_1}},~
\widetilde{K}(\beta)=\lim_{\rho \to \infty}
K(\beta).
\end{eqnarray}
The matrix $\widetilde{S}(\beta)$ and
the matrix $\widetilde{K}(\beta)$ satisfy Yang-Baxter
equation and the Boundary Yang-Baxter equation, too. 
The pair $\widetilde{K}(\beta)$ and $\widetilde{S}(\beta)$
exist in the class which Hou et al. studied \cite{HSY},
because
the matrix $\widetilde{S}(\beta)$
satisfies the projection property,
$
\widetilde{S}\left(\frac{2\pi i}{n}\right) \sim P_2^{(-)},
$
where $P_2^{(-)}$ is antisymmetric project operator
in $V \otimes V$.
The matrix $\widetilde{S}(\beta)$ and
the $K$-matrix $\widetilde{K}(\beta)$ satisfy 
boundary crossing symmetry. See notation in
reference \cite{HSY}.
\begin{eqnarray}
\widetilde{K}^*(\beta)_j^k=
\sum_{l,m=0}^{n-1}
\widetilde{S}(2\beta-2\pi i)_{k,m}^{l,j}
\widetilde{K}(-\beta+2\pi i)_{l}^{m}.
\label{BCS}
\end{eqnarray}
For example,
$n=3$ case of the relation (\ref{BCS})
in a limiting case
$\rho \to \infty$, becomes 
\begin{eqnarray}
\varphi\left(\beta-\frac{2\pi i}{3}\right)
\varphi(\beta)
\varphi\left(\beta+\frac{2\pi i}{3}\right)=
\frac{S_1(-2i\beta-\frac{4\pi}{3}|\rho)}
{S_1(2i\beta-\frac{4\pi}{3}|\rho)
}
\rightarrow
\frac{-\beta+\frac{2\pi i}{3}}{
\beta+\frac{2\pi i}{3}}.
\end{eqnarray}

For the description of the space of the physical states,
we use the Zamolodchikov-Faddeev operators.
The Zamolodchikov-Faddeev operators
$Z_j^*(\beta), Z_j(\beta), (j=0,\cdots, n-1)$
of the present model are characterized by the
following three conditions.
\begin{eqnarray}
Z_{j_1}^*(\beta_1)Z_{j_2}^*(\beta_2)
&=&\sum_{k_1,k_2=0}^{n-1}
Z_{k_2}^*(\beta_2)Z_{k_1}^*(\beta_1)
S(\beta_1-\beta_2)_{j_1,j_2}^{k_1,k_2},
\label{Com1}\\
Z_{j_1}(\beta_1)Z_{j_2}(\beta_2)
&=&\sum_{k_1,k_2=0}^{n-1}
S(\beta_1-\beta_2)_{k_1,k_2}^{j_1,j_2}
Z_{k_2}(\beta_2)Z_{k_1}(\beta_1),\label{Com2}
\end{eqnarray}
The Zamolodchikov-Faddeev operators $
Z_j(\beta), Z_j^*(\beta)
$ satisfy the inversion relation.
\begin{eqnarray}
Z_j^*(\beta_1)
Z_k(\beta_2+\pi i)=
\frac{\delta_{j,k}}{\beta_1-\beta_2}+\cdots,~(\beta_1
\to \beta_2),
\label{Inversion}
\end{eqnarray}
where $''\cdots''$ means regular term.
Free field realizations of the
Zamolodchikov-Faddeev operators were given in \cite{MT}.

In terminology of quantum field theory,
the operator ${\cal O}$
which commutes with the Zamolodchikov-Faddeev
operators up to scalar multiplicity,
is called the local operator. 
\begin{eqnarray}
Z_j^*(\beta){\cal O}=m(\beta)
{\cal O}
Z_j^*(\beta).
\end{eqnarray}
In this paper we restrict our attention to
a class of local operators $Z_j'(\delta), (j=0,\cdots,n-1)$
which commuts with the Zamolodchikov-Faddeev operators
as follows.
\begin{eqnarray}
Z_j^*(\beta)Z_k'(\delta)=
{\cal L}(\beta-\delta)Z_k'(\delta)Z_j^*(\beta),~~
{\cal L}(\delta)=
\frac{{\rm sh}(-\delta+\frac{\pi i}{n})}
{{\rm sh}(\delta+\frac{\pi i}{n})}.
\end{eqnarray}
Free field realizations of a class of local operators
$Z_k'(\delta),~(k=0,\cdots,n-1)$ were given in
the reference \cite{KY},
in which the authors considered spin chain problem.

For the description of the space of the physical state
we use the boundary state $|B\rangle$.
The boundary state $|B\rangle$
and its dual $\langle B|$
are characterized by the following conditions.
\begin{eqnarray}
K(\beta)_j^j Z_j^*(\beta)| B \rangle &=& Z_j^*(-\beta)
|B \rangle,~(j=0,\cdots,n-1),\\
K(\beta)_j^j\langle B| Z_j(-\beta+\pi i)&=&
\langle B| Z_j(\beta+\pi i),~
(j=0,\cdots,n-1).
\end{eqnarray}
In this paper we shall construct the free field
realizations of the boundary state and its dual.
The space of state is spanned by the vectors,
\begin{eqnarray}
Z_{j_1}^*(\beta_1)\cdots Z_{j_N}^*(\beta_N)|B\rangle.
\end{eqnarray}

Let us set the local operator ${\cal Z}$ by
\begin{eqnarray}
{\cal Z}=Z_{k_1}'(\delta_1)\cdots Z_{k_Q}'(\delta_Q).
\end{eqnarray}
Consider the matrix element,
\begin{eqnarray}
\langle B |{\cal Z}
Z_{j_1}^*(\beta_1)\cdots Z_{j_N}^*(\beta_N)|B\rangle.
\end{eqnarray}
We call the above matrix elements
``form factors''.
In this paper we give integral representations
for form factors of local fields.

In this section we have introduced the basic tools
of boundary field theory, i.e.
the $S$-matrix $S(\beta)$,
the boundary $K$-matrix $K(\beta)$,
the Zamolodchikov-Faddeev operators $Z_j^*(\beta)$,
local operator ${\cal Z}$,
the boundary state $|B\rangle$, 
its dual $\langle B|$, and form factors.

\section{ZF operators,
Local operators}
The puropose of this section
is to give the free field realizations of
Zamolodchikov-Faddeev operators
\cite{MT} and local operators \cite{KY}
of the present model.

\subsection{ZF operators}

The purpose of this subsection is
to give free field realizations of
the Zamolodchikov-Faddeev operators
$Z_j^*(\beta), Z_j(\beta), (j=0,\cdots,n-1)$.
Let us introduce the free bosons,
$a_j(t) (j=1,\cdots,n-1 ; t \in {\mathbb{R}})$,
satisfying the commutation relations,
\begin{eqnarray}
[a_j(t),a_k(t')]=A_{j k}(t)\delta(t+t'),~
A_{j k}(t)=-\frac{1}{t}
\frac{{\rm sh}\frac{(a_j|a_k)\pi t}{n}
{\rm sh}\left(\frac{\rho}{2}+\frac{\pi}{n}\right)t
}{
{\rm sh}\frac{\pi t}{n}
{\rm sh}\frac{\rho t}{2}
}.
\end{eqnarray}
Here $((a_j|a_k))_{1\leq j,k \leq n-1}$
is the Cartan matrix of type $A_{n-1}$.
Explicitly the Cartan matrix
of type $A_{n-1}$, $((a_j|a_k))_{1\leq j,k \leq n-1}$,
is written as
\begin{eqnarray}
\left(\begin{array}{cccccc}
2&-1&0&\cdots&\cdots&0\\
-1&2&-1&0&\cdots&0\\
0&-1&2&-1&0&\cdots\\
\cdots&\cdots&\cdots&\cdots&\cdots&\cdots\\
\cdots&\cdots&0&-1&2&-1\\
0&\cdots&\cdots&0&-1&2
\end{array}\right)
\end{eqnarray}
Let us introduce the Fock space ${\cal F}$
generated by the vacuum vector $|vac\rangle$
satisfying
\begin{eqnarray}
a_j(t)|vac\rangle=0~{\rm for}~t>0. 
\end{eqnarray}
A normal ordering $:A:$ of an element $A$
is defined as usual ;
annihilation operators $a_j(t) (t>0)$
are replaced on the right of the creation operators
$a_j(-t) (t>0)$,
for example,
\begin{eqnarray}
:a_j(t_1)a_k(-t_2):=
a_k(-t_2)a_j(t_1), (t_1, t_2>0).
\end{eqnarray}

Let us introduce the basic operators $V_j(\alpha)
(j=0,1,\cdots, n)$ by
\begin{eqnarray}
V_j(\alpha)&=&:\exp\left(\int_{-\infty}
^\infty
a_j(t)e^{i \alpha t} dt\right):
(j=1,\cdots, n-1),\\
V_0(\alpha)&=&:\exp\left(\int_{-\infty}
^\infty
a_1^*(t)e^{i \alpha t} dt\right):,\\
V_{n}(\alpha)&=&:\exp\left(\int_{-\infty}
^\infty
a_{n-1}^*(t)e^{i \alpha t} dt\right):,
\end{eqnarray}
where
\begin{eqnarray}
a_1^*(t)=-\sum_{j=1}^{n-1}
a_j(t)\frac{{\rm sh}\frac{(n-j)\pi t}{n}}{
{\rm sh}\pi t
},~
a_{n-1}^*(t)=-\sum_{j=1}^{n-1}
a_j(t)\frac{{\rm sh}\frac{j \pi t}{n}}{
{\rm sh}\pi t
}.
\end{eqnarray}
We have
\begin{eqnarray}
[a_1^*(t),a_j(t')]=\delta_{1,j}\frac{1}{t}
\frac{{\rm sh}\left(\frac{\rho}{2}+
\frac{\pi}{n}\right)t}{{\rm sh}\frac{\rho t}{2}}\delta(t+t')
,~~
[a_j(t),a_{n-1}^*(t')]=\delta_{j,n-1}\frac{1}{t}
\frac{{\rm sh}\left(\frac{\rho}{2}+
\frac{\pi}{n}\right)t}{{\rm sh}\frac{\rho t}{2}}
\delta(t+t').\nonumber\\
\end{eqnarray}

Free field realizations of the Zamolodchikov-Faddeev
operators
$Z_j(\beta)$,
of the affine $A_{n-1}^{(1)}$ Toda fields
are given by

\begin{eqnarray}
Z_j(\beta)&=&c_j
\int_{C_{j+1}} d\alpha_{j+1}\cdots \int_{C_{n-1}}
d\alpha_{n-1}
:V_{j+1}(\alpha_{j+1})\cdots
V_{n-1}(\alpha_{n-1})V_n(\beta):\nonumber
\\
&\times&
e^{\frac{\pi}{\rho}(\alpha_{j+1}-\beta)}
\prod_{k=j+1}^{n-1}
\Gamma\left(\frac{i(\alpha_{k+1}-\alpha_k)}{\rho}
-\frac{\pi}{n \rho}\right)
\Gamma\left(\frac{i(\alpha_{k}-\alpha_{k+1})}{\rho}
-\frac{\pi}{n \rho}\right),\label{real:Psi}
\end{eqnarray}
where $\beta=\alpha_n$.
Here the integral contour $C_k, (k=1,\cdots,n-1)$
for $\alpha_k$ is chosen so that
the poles at $\alpha_{k+1}-\frac{\pi i}{n}+\rho i 
{\mathbb{Z}}_{\geq 0}$ are above $C_k$, and
that
the poles at $\alpha_{k+1}+\frac{\pi i}{n}-\rho i
{\mathbb{Z}}_{\geq 0}$ are below $C_k$.
Here constant $c_j$ is chosen so that
the inversion relation (\ref{Inversion}) holds.

Free fields realizations of the Zamolodchikov-Faddeev
operators $Z_j^*(\beta)$ are given by
\begin{eqnarray}
Z_j^*(\beta)&=&\int_{C_1^*} d\alpha_1
\cdots \int_{C_j^*} d\alpha_j
:V_0(\beta)V_1(\alpha_1)\cdots V_j(\alpha_j):\nonumber\\
&\times&
e^{\frac{\pi}{\rho}(\beta-\alpha_j)}
\prod_{k=1}^j
\Gamma\left(\frac{i(\alpha_{k-1}-\alpha_k)}{\rho}
-\frac{\pi}{n \rho}\right)
\Gamma\left(\frac{i(\alpha_{k}-\alpha_{k-1})}{\rho}
-\frac{\pi}{n \rho}\right),
\label{real:Psi*}
\end{eqnarray}
where $\beta=\alpha_0$.
Here
the integral contour $C_{k}^*, (k=1,\cdots,n-1)$
for $\alpha_k$ is chosen so that
the poles at $\alpha_{k-1}-\frac{\pi i}{n}+\rho i 
{\mathbb{Z}}_{\geq 0}$ are above $C_k^*$, and
that
the poles at $\alpha_{k-1}+\frac{\pi i}{n}-\rho i
{\mathbb{Z}}_{\geq 0}$ are below $C_k^*$.

Proof of the commuation relation (\ref{Com1})
was given in \cite{MT}.
The commutation relation (\ref{Com2})
is derived as the same manner as (\ref{Com1}).
The inversion relation (\ref{Inversion})
is derived as the same manner as those of
local operators in \cite{KY}.

In what follows we use the following abberiviations,
$(j=1,\cdots,n-1)$.
\begin{eqnarray}
V_j^a(\alpha)=\exp\left(\int_0^\infty a_j(t)
e^{i\alpha t}dt\right),~
V_j^c(\alpha)=\exp\left(
\int_0^\infty a_j(-t)
e^{-i\alpha t}dt\right),\\
V_0^a(\alpha)=\exp\left(\int_0^\infty a_1^*(t)
e^{i\alpha t}dt\right),~
V_0^c(\alpha)=\exp\left(
\int_0^\infty a_1^*(-t)
e^{-i\alpha t}dt\right),\\
V_{n}^a(\alpha)=\exp\left(\int_0^\infty a_{n-1}^*(t)
e^{i\alpha t}dt\right),~
V_n^c(\alpha)=\exp\left(
\int_0^\infty a_{n-1}^*(-t)
e^{-i\alpha t}dt\right).
\end{eqnarray}

~\\
{\it Note.~~
Free field realizations of
the Zamolodchikov-Faddeev operators $Z_j^*(\beta)$
were given in \cite{MT}.
In this paper we give free field realizations
of the dual operators
$Z_j(\beta)$.
}

\subsection{Local operators}

In terminology of quantum field theory,
the local operator is the one
which commutes with the Zamolodchikov-Faddeev operators,
up to scalar function multiplicity.
In this subsection we present free field realization
of a class of local operators $Z_j'(\beta)$
constructed in \cite{KY}.
The local operators $Z_j'(\beta), (j=0,\cdots,n-1)$
commute with the Zamolodchikov-Faddeev operators,
up to multiplicity function.
\begin{eqnarray}
Z_j^*(\beta)Z_k'(\delta)=
{\cal L}(\beta-\delta)Z_k'(\delta)Z_j^*(\beta),
\end{eqnarray}
where we have set
\begin{eqnarray}
{\cal L}(\beta)=
\frac{{\rm sh}(-\beta+\frac{\pi i}{n})}
{{\rm sh}(\beta+\frac{\pi i}{n})}.
\end{eqnarray}
Let ue set the auxiliary fields $b_j(t),
(1\leq j \leq n-1; t \in {\mathbb{R}})$
by
\begin{eqnarray}
b_j(t)=\frac{{\rm sh}\frac{\rho t}{2}}{
{\rm sh}\left(\frac{\rho}{2}+
\frac{\pi}{n}\right)t}\times a_j(t).
\end{eqnarray}
The bose field $b_j(t)$ satisfies the following
commutation relation.
\begin{eqnarray}
[b_j(t),b_k(t')]=-\frac{1}{t}
\frac{{\rm sh}\frac{(a_j|a_k)\pi t}{n}
{\rm sh}\frac{\rho t}{2}
}{
{\rm sh}\frac{\pi t}{n}
{\rm sh}\left(\frac{\rho}{2}+\frac{\pi}{n}\right)t
}\delta(t+t').
\end{eqnarray}
Let us set
\begin{eqnarray}
b_1^*(t)=-\sum_{j=1}^{n-1}b_j(t)
\frac{{\rm sh}\frac{(n-j)\pi t}{n}}{{\rm sh}\pi t}.
\end{eqnarray}
We have
\begin{eqnarray}
[b_1^*(t),b_j(t')]=\delta_{j,1}\frac{1}{t}
\frac{
{\rm sh}\frac{\rho t}{2}
}{
{\rm sh}\left(\frac{\rho}{2}+\frac{\pi}{n}\right)t
}\delta(t+t').
\end{eqnarray}
Let us set the basic operators.
\begin{eqnarray}
U_j(\delta)&=&:\exp\left(-\int_{-\infty}
^\infty b_j(t)e^{i\delta t}\right):, (1\leq j \leq n-1),\\
U_0(\delta)&=&:\exp\left(-\int_{-\infty}^\infty
b_1^*(t)e^{i \delta t}dt\right):.
\end{eqnarray}
Free field realizations of the local operators
$Z_j'(\delta), (j=0,\cdots,n-1)$ are given by
\begin{eqnarray}
Z_j'(\delta)&=&
\int_{-\infty}^\infty d\gamma_1 \cdots
\int_{-\infty}^\infty d\gamma_j 
:U_0(\delta)U_1(\gamma_1)\cdots U_j(\gamma_j):\nonumber
\\
&\times&
e^{\frac{\pi}{\rho+\frac{2\pi}{n}}(\gamma_j-\delta)}
\prod_{k=1}^j
\Gamma\left(
\frac{i(\gamma_k-\gamma_{k-1})+\frac{\pi}{n}}{
\rho+\frac{2\pi}{n}}
\right)
\Gamma\left(
\frac{i(\gamma_{k-1}-\gamma_{k})+\frac{\pi}{n}}{
\rho+\frac{2\pi}{n}}
\right).
\end{eqnarray}
Here we set $\gamma_0=\delta$.

~\\
{\it Note.~
Free field realizations of operators $Z_j'(\delta)$
were given in the reference \cite{KY},
in which authors considered correlation functions of
the critical $A_{n-1}^{(1)}$ chain.}

\section{Boundary state}
The purpose of this section is
to give the free field realizations
of the boundary state $|B\rangle$ and its dual
$\langle B|$.
We give free field realizations of
the boundary state $|B\rangle$,
\begin{eqnarray}
K(\beta)_j^j
Z_j^*(\beta)|B\rangle=
Z_j^*(-\beta)|B\rangle,~(j=0,\cdots,n-1),
\label{Ref1}
\end{eqnarray}
and its dual state $\langle B|$,
\begin{eqnarray}
K(\beta)_j^j
\langle B|Z_j(-\beta+\lambda i)=
\langle B|Z_j(\beta+\lambda i),~
(j=0,\cdots,n-1; \lambda>0).
\label{Ref2}
\end{eqnarray}
Here we have used $K(\beta)_j^j$ given in (\ref{def:K}).
When we set $\lambda=\pi$,
dual state $\langle B|$ becomes the dual boundary state
of the affine $A_{n-1}^{(1)}$ Toda fields.
When we set
$\lambda=0$,
vacuum expectation value,
$\langle B |Z_{j_1}^*(\beta_1)
\cdots Z_{j_N}^*(\beta_N)|B\rangle
$ 
produces 
an eigenvector of 
$A_{n-1}^{(1)}$ analogue
of finite XXZ chain with double boundaries.
See reference \cite{FW}.

\subsection{Boundary state}

In this subsection we give the free field realization
of the boundary state $|B\rangle$, and
show the reflection realtion (\ref{Ref1}).
\\
{\it The boundary state $|B\rangle$
is realized as follows.}
\begin{eqnarray}
|B\rangle=e^{B}|vac\rangle.\label{boundary1}
\end{eqnarray}
{\it Here $B$ is a quadratic part of free bosons.}
\begin{eqnarray}
B=\sum_{j,k=1}^{n-1}\int_0^\infty
\alpha_{j,k}(t)a_j(-t)a_k(-t)dt
+\sum_{j=1}^{n-1}\int_0^\infty
\beta_j(t)a_j(-t)dt,
\end{eqnarray}
{\it where scalars $\alpha_{j,k}(t) (j,k=1,\cdots,n-1)$ 
and $\beta_j(t) (j=1,\cdots, n-1)$
are given by}
\begin{eqnarray}
\alpha_{j,k}(t)&=&
-\frac{t}{2}\times
\frac{\displaystyle
{\rm sh}\left(\frac{\rho t}{2}\right)}{
\displaystyle
{\rm sh}\left(
\left(\frac{\rho}{2}+\frac{\pi}{n}\right)t\right)
}\times I_{j,k}(t),\label{alpha}
\\
\beta_j(t)
&=&-
\frac{
\displaystyle
{\rm ch}\left(
\frac{\rho t}{4}\right)}{
\displaystyle
{\rm sh}
\left(\left(\frac{\rho}{4}+\frac{\pi}{2n}\right)t
\right)}\times
\frac{\displaystyle
{\rm sh}^2\left(\frac{\pi t}{2}\right)
}{\displaystyle
{\rm sh}(\pi t)
}
\times
I_{j,j}\left(
\frac{t}{2}\right)
\nonumber\\
&=&-
{\rm sh}\left(\frac{\pi t}{2n}\right)\times
\frac{
\displaystyle
{\rm ch}\left(
\frac{\rho t}{4}\right)}{
\displaystyle
{\rm sh}
\left(\left(\frac{\rho}{4}+\frac{\pi}{2n}\right)t
\right)}\times
\sum_{k=1}^{n-1}I_{k,j}(t).\label{beta}
\end{eqnarray}
{\it Here we have set
the symmetric matrix $\left(I_{j,k}(t)\right)
_{1\leq j,k \leq n-1}$ by}
\begin{eqnarray}
I_{j,k}(t)=\frac{
\displaystyle
{\rm sh}\left(\frac{j\pi t}{n}\right)
{\rm sh}\left(\frac{(n-k)\pi t}{n}\right)
}{\displaystyle
{\rm sh}\left(\frac{\pi t}{n}\right)
{\rm sh}\left(\pi t\right)
}=I_{k,j}(t),
~~(1\leq j \leq k \leq n-1).
\label{matrix:I}
\end{eqnarray}
{\it We remark that
the matrix} $(I_{j,k}(t))_{1\leq j,k \leq n-1}$
{\it is the inverse matrix of}
$\left(\frac{{\rm sh}\frac{(a_j|a_k)\pi t}{n}}
{{\rm sh}\frac{\pi t}{n}}\right)_{1\leq j,k \leq n-1}$.

Let us show that
the boundary state $|B\rangle$
satisfies the following reflection
relations.
\begin{eqnarray}
Z_j^*(\beta)|B\rangle=
\varphi(\beta)
Z_j^*(-\beta)|B\rangle.
\end{eqnarray}
The presence of $e^{B}$
has an effect of a Bogoliubov transformation.
\begin{eqnarray}
e^{-B}a_l(t)e^{B}&=&a_l(t)+a_l(-t)+\sum_{j=1}^{n-1}
A_{lj}(t)\beta_j(t),~~(t \geq 0)\nonumber\\
&=&a_l(t)+a_l(-t)+\frac{1}{t}
\frac{
\displaystyle
{\rm ch}\left(\left(\frac{\rho}{4}+
\frac{\pi}{2n}\right)t\right)}{
\displaystyle
{\rm sh}
\left(\frac{\rho t}{4}\right)} 
~{\rm sh}\left(\frac{\pi t}{2n}\right)
,\\
e^{-B}a_l(-t)e^{B}&=&a_l(-t),~~(t>0).
\end{eqnarray}
Therefore the basic operators $V_j(\alpha)$
act on the boundary state $|B\rangle$ as follows.
\begin{eqnarray}
V_j^a(\alpha)|B\rangle=G_j(\alpha)
V_j^c(-\alpha)|B\rangle,~(j=0,1,\cdots,n).
\end{eqnarray}
We have set
\begin{eqnarray}
G_j(\alpha)&=&
-i2^{1+\frac{2\pi}{n\rho}}e^\gamma
\times \alpha, ~~(j=1,\cdots,n-1),
\label{Gj}\\
G_0(\alpha)&=&
2^{-\frac{\pi}{\rho}\frac{n-1}{n}}
e^{\gamma \frac{n-1}{2n}(1+\frac{2\pi(n+1)}{\rho n})}
\nonumber\\
&\times&
\frac{\Gamma_2\left(\left.
-2i\alpha+2\pi-\frac{2\pi}{n}\right|\rho,4\pi\right)
\Gamma_2\left(\left.
-2i\alpha+\rho+2\pi+\frac{2\pi}{n}\right|\rho,4\pi\right)
}{
\Gamma_2\left(\left.
-2i\alpha \right|\rho,4\pi\right)
\Gamma_2\left(\left.
-2i\alpha+\rho+4\pi \right|\rho,4\pi\right)
},
\label{G0}\\
G_n(\alpha)&=&G_0(\alpha).
\label{Gn}
\end{eqnarray}
Using the above formuale of the action of the
basic operators, we get the actions of
the Vertex opertaors $Z_j^*(\beta)$
as follows.
\begin{eqnarray}
Z_j^*(\beta)|B\rangle&=&
(2\pi)^{-j}
(\rho e^\gamma e^{-\frac{\pi i}{2}})^{-(1+\frac{2\pi}{n\rho})j}
\int_{C_1^*} d\alpha_1 \cdots \int_{C_j^*} d\alpha_j
\prod_{k=0}^j V_k^c(\alpha_k)V_k^c(-\alpha_k)|B\rangle
\nonumber\\
&\times&
\prod_{k=0}^j G_k(\alpha_k)
\prod_{k=1}^j \Delta(\alpha_{k-1},\alpha_k)
\prod_{k=1}^j S(\alpha_{k-1},\alpha_k).
\end{eqnarray}
Here we have set the auxiliary functions
$\Delta(\alpha_1,\alpha_2)$ and $S(\alpha_1, \alpha_2)$
by
\begin{eqnarray}
\Delta(\alpha_1,\alpha_2)&=&
\prod_{\epsilon_1, \epsilon_2=\pm}
\Gamma\left(
\frac{i(\epsilon_1 \alpha_1+\epsilon_2 \alpha_2)}{\rho}
-\frac{\pi}{n \rho}
\right),\label{fun:Delta}
\\
S(\alpha_1,\alpha_2)&=&e^{\frac{2\pi}{\rho}\alpha_1}
-q e^{-\frac{2\pi}{\rho}\alpha_2},
~q=e^{-\frac{2\pi^2 i}{\rho n}}
.\label{fun:S}
\end{eqnarray}
The auxiliary function $\Delta(\alpha_1,\alpha_2)$
is invariant under the change of variables 
$\alpha_j \to -\alpha_j$,
\begin{eqnarray}
\Delta(\alpha_1,\alpha_2)=
\Delta(-\alpha_1,\alpha_2)=
\Delta(\alpha_1,-\alpha_2)=
\Delta(-\alpha_1,-\alpha_2).
\end{eqnarray}
Now we arrive at
\begin{eqnarray}
&&Z_j^*(\beta)|B\rangle-
\varphi(\beta)
Z_j^*(-\beta)|B\rangle \nonumber\\
&=& 2^{-j+1}
\pi^{-j}
(\rho e^\gamma e^{-\frac{\pi i}{2}})^{-(1+\frac{2\pi}{n\rho})j}
\times {\rm sh}\left(\frac{2\pi}{\rho}\beta\right)
G_0(\beta)\int_{C_1^*} d\alpha_1 \cdots 
\int_{C_j^*} d\alpha_j
\nonumber\\
&\times&
\prod_{k=0}^j
V_k^c(\alpha_k)V_k^c(-\alpha_k)|B\rangle
\prod_{k=1}^j
\Delta(\alpha_{k-1},\alpha_k)\prod_{k=2}^j
S(\alpha_{k-1},\alpha_k)\prod_{k=1}^j G_k(\alpha_k).
\end{eqnarray}
Here we have used the relation,
\begin{eqnarray}
\varphi(\beta)=\frac{G_0(\beta)}{
G_0(-\beta)}.
\end{eqnarray}
We change the integral variable $\alpha_k \leftrightarrow
-\alpha_k$, and the corresponding new contour
$\widetilde{C}^*_k$.
We find that the corresponding new contour $
\widetilde{C}^*_k
$ can be deformed to the same as $C_k^*$.
We find that the following part in the integrand :
\begin{eqnarray}
\prod_{k=1}^j \Delta(\alpha_{k-1},\alpha_k)
\prod_{k=0}^j V_k^c(\alpha_k) V_k^c(-\alpha_k)|B\rangle,
\end{eqnarray}
is invariant under the change of variable
$\alpha_k \leftrightarrow -\alpha_k$.
Summning up all changing of integral variables
$\pm \alpha_k, (k=1,\cdots,j)$,
we know that
a sufficient condition
of the relation,
\begin{eqnarray}
Z_j^*(\beta)|B\rangle-\varphi(\beta)
Z_j^*(-\beta)|B\rangle=0,
\end{eqnarray}
becomes the following polynomial identity,
\begin{eqnarray}
\sum_{\epsilon_1,\cdots,\epsilon_j=\pm}
\prod_{k=2}^j S(\epsilon_{k-1}\alpha_{k-1},
\epsilon_k \alpha_k)
\prod_{k=1}^j G_k(\epsilon_k \alpha_k)=0.
\end{eqnarray}
Now we have derived the reflection
relation (\ref{Ref1}).

\subsection{Dual Boundary State}

{\it The dual boundary state $\langle B|$
is realized as follows.}
\begin{eqnarray}
\langle B|=\langle vac |e^G.\label{boundary2}
\end{eqnarray}
{\it Here $G$ is a quadratic form of free bosons.}
\begin{eqnarray}
G=\sum_{j,k=1}^{n-1}
\int_0^\infty \gamma_{j,k}(t)
a_j(t)a_k(t)dt+
\sum_{j=1}^{n-1}\int_0^\infty \delta_j(t)a_j(t)dt,
\end{eqnarray}
{\it where scalar $\gamma_{j,k}(t) (j,k=1,\cdots,n-1)$
and $\delta_j(t) (j=1, \cdots, n-1)$
are given by}
\begin{eqnarray}
\gamma_{j,k}(t)=
e^{-2\lambda t}\alpha_{j,k}(t),~~~
\delta_j(t)=e^{-\lambda t}\beta_j(t).
\end{eqnarray}
{\it Here $\alpha_{j,k}(t)$ and
$\beta_j(t)$
are introduced in (\ref{alpha}) and (\ref{beta}).}

Let us show the dual boundary state $\langle B|$
satisfies the following reflection relations.
\begin{eqnarray}
\langle B|
Z_j(-\beta+\lambda i)=
\varphi(\beta)
\langle B|Z_j(\beta+\lambda i), ~(j=0,\cdots,n-1).
\end{eqnarray}
The precence of $e^G$ has an effect of a Bogoliubov
transformation.
\begin{eqnarray}
e^G a_l(-t) e^{-G}&=&a_l(-t)+e^{-2 \lambda t}a_l(t)+
\sum_{j=1}^{n-1}A_{l,j}(t)\delta_j(t), (t\geq0)
\nonumber\\
&=&a_l(-t)+e^{-2 \lambda t}a_l(t)+
\frac{e^{-\lambda t}}{t}
\frac{\displaystyle
{\rm ch}\left(
\left(\frac{\rho}{4}+\frac{\pi}{2n}\right)t
\right)}{\displaystyle
{\rm sh}
\left(\frac{\rho t}{4}\right)
}
{\rm sh}\left(\frac{\pi t}{2n}\right),
\\
e^G a_l(t) e^{-G}&=&a_l(t), ~~~(t>0).
\end{eqnarray}
Therefore
the basic operators $V_j(\alpha)$ act on
the dual boundary state $\langle B|$ as follows.
\begin{eqnarray}
\langle B|V_j^c(\alpha+\lambda i)&=&
G_j^*(\alpha)\langle B|V_j^a(-\alpha+\lambda i),~
(j=0, \cdots, n).
\end{eqnarray}
We have set
\begin{eqnarray}
&&G_j^*(\alpha)=G_j(-\alpha),~(j=0,\cdots,n).
\end{eqnarray}
Here $G_j(\alpha)$ is defined in
(\ref{Gj}), (\ref{G0}) and (\ref{Gn}).\\
We get the actions of the Vertex operators 
$Z_j(\beta)$
as follows.
\begin{eqnarray}
\langle B|Z_j(\beta+\lambda i)&=&
c_j (2\pi)^{-(n-j-1)}
(\rho e^\gamma e^{-\frac{\pi i}{2}})^{-(1+\frac{2\pi}{\rho n})
(n-1-j)}
\int_{C_{j+1}} d\alpha_{j+1} 
\cdots \int_{C_{n-1}} d\alpha_{n-1}
\nonumber\\
&\times&
\langle B|
\prod_{k=j+1}^{n}
V_k^a(\alpha_k+\lambda i)
V_k^a(-\alpha_k+\lambda i)
\nonumber\\
&\times&
\prod_{k=j+1}^n G_k^*(\alpha_k)
\prod_{k=j+1}^{n-1}
\Delta(\alpha_k,\alpha_{k+1})
\prod_{k=j+1}^{n-1}
S(\alpha_k,\alpha_{k+1}),
\end{eqnarray}
where $\Delta(\alpha_1, \alpha_2)$
and $S(\alpha_1, \alpha_2)$
are given in (\ref{fun:Delta}) and (\ref{fun:S}).
Now we have
\begin{eqnarray}
&&\langle B|Z_j(\beta+\lambda i)-
\varphi(\beta)^{-1}
\langle B|Z_j(-\beta+\lambda i)\nonumber\\
&=&
2^{-n+j+2}c_j
e^{-\frac{2\pi^2 i}{\rho n}}
\pi^{-n+j+1}(\rho e^\gamma e^{-\frac{\pi}{2}})
^{-(1+\frac{2\pi}{\rho n})(n-1-j)}
{\rm sh}\left(\frac{2\pi}{\rho} \beta \right)
G_n^*(\beta)\nonumber\\
&\times&
\int_{C_{j+1}} d\alpha_{j+1} \cdots 
\int_{C_{n-1}} d\alpha_{n-1}
\langle B|
\prod_{k=j+1}^{n}
V_k^a(\alpha_k+\lambda i)
V_k^a(-\alpha_k+\lambda i)\nonumber\\
&\times&
\prod_{k=j+1}^{n-1}\Delta(\alpha_k, \alpha_{k+1})
\prod_{k=j+1}^{n-2}S(\alpha_k,\alpha_{k+1})
\prod_{k=j+1}^{n-1}G_k^*(\alpha_k).
\end{eqnarray}
As the same arguments as the case of the boundary state,
we get a sufficient condition of
the reflection relation (\ref{Ref2}),
\begin{eqnarray}
\sum_{\epsilon_{j+1} \cdots \epsilon_{n-1}=\pm}
\prod_{k=j+1}^{n-2}
S(\epsilon_k \alpha_k, \epsilon_{k+1} \alpha_{k+1})
\prod_{k=j+1}^{n-1}G_k^*(\epsilon_k \alpha_k)=0.
\end{eqnarray}
Now we have derived the reflection relation
(\ref{Ref2}).

In this section,
we have constructed free field realizations
of the boundary state $|B\rangle$
and its dual $\langle B|$.

\section{Form factors}

The purpose of this section is
to derive the difference equations
which govern the form factors, and
to give integral representations
for form factors. 

Let us introduce matrix element, called form factors.
\begin{eqnarray}
f^{{\cal Z}}(\beta_1,\cdots,\beta_N)
_{j_1 \cdots j_N}\nonumber=
\frac{\langle B|
{\cal Z}
Z_{j_1}^*(\beta_1)\cdots Z_{j_N}^*(\beta_N)
|B\rangle}{
\langle B|B \rangle}.
\end{eqnarray}
Here $Z_j^*(\beta)$ is the Zamolodchikov-Faddeev operators,
$|B\rangle$ is the boundary state,
and $
{\cal Z}=Z_{k_1}'(\delta_1)\cdots Z_{k_Q}'(\delta_Q),
~(k_1,\cdots,k_Q=0,\cdots,n-1)$.
Let us set
\begin{eqnarray}
f^{{\cal Z}}(\beta_1,\cdots,\beta_N)
=\sum_{j_1,\cdots,j_N=0}^{n-1}
v_{j_1}\otimes \cdots \otimes v_{j_N}
f^{{\cal Z}}(\beta_1,\cdots,\beta_N)_{j_1
\cdots j_N}.
\end{eqnarray}

\subsection{Boundary qKZ equations}

The purpose of this section is to derive
the difference equations 
which govern form factors.
Let us introduce notation
$Z^*(\beta), Z^{*(1)}(\beta_1)Z^{*(2)}(\beta_2)$
and $Z^{*(2)}(\beta_2)Z^{*(1)}(\beta_1)
$ as follows.
\begin{eqnarray}
Z^{*}(\beta)&=&\sum_{j=0}^{n-1}
Z^*_j(\beta)\otimes v_j,\\
Z^{*(1)}(\beta_1)Z^{*(2)}(\beta_2)
&=&\sum_{j_1,j_2=0}^{n-1}
Z^*_{j_1}(\beta_1)Z^*_{j_2}(\beta_2)
\otimes v_{j_1} \otimes v_{j_2},\\
Z^{*(2)}(\beta_2)Z^{*(1)}(\beta_1)
&=&\sum_{j_1,j_2=0}^{n-1}
Z^*_{j_2}(\beta_2)Z^*_{j_1}(\beta_1)
\otimes v_{j_1} \otimes v_{j_2}.
\end{eqnarray}
In this notation
the commutation relation of the Zamolodchikov-Faddeev
operators is written by
\begin{eqnarray}
Z^{*(1)}(\beta_1)Z^{*(2)}(\beta_2)=
S_{12}(\beta_1-\beta_2)
Z^{*(2)}(\beta_2)Z^{*(1)}(\beta_1),
\end{eqnarray}
where the $S$-matrix $S_{12}(\beta)$
acts on the space $V \otimes V$.
As the same manner we write the reflection relations
(\ref{Ref1}), (\ref{Ref2})
as follows.
\begin{eqnarray}
Z^*(\beta)|B\rangle&=&K(-\beta)Z^*(-\beta)|B\rangle,\\
\langle B|Z(\beta+i\lambda)&=&
K(\beta)\langle B|Z(-\beta+i\lambda).
\end{eqnarray}
As the same manner as (\ref{Ref1}), (\ref{Ref2}),
we have the following reflection relations.
\begin{eqnarray}
\langle B|Z^*(\beta+i\lambda)&=&
K(\beta)\langle B|Z^*(-\beta+i\lambda),\\
Z(\beta)|B\rangle&=&K(-\beta)Z(-\beta)|B\rangle.
\end{eqnarray}
In this notation the function $f^{{\cal Z}}(\beta_1,
\cdots,\beta_N)$ is written as
\begin{eqnarray}
f^{{\cal Z}}(\beta_1,
\cdots,\beta_N)=
\frac{
\langle B|{\cal Z}Z^{*(1)}(\beta_1)
\cdots Z^{*(N)}(\beta_N)|B\rangle}{
\langle B|B \rangle}.
\end{eqnarray}

Let us derive the difference equations,
which govern the form factor $f^{{\cal Z}}(\beta_1,\cdots,
\beta_N)$.
Consider the vacuum expectation value,
\begin{eqnarray}
\langle B|{\cal Z}Z^{*(1)}(\beta_1)
\cdots Z^{*(r)}(\beta_r) \cdots Z^{*(N)}(\beta_N)|B\rangle.
\end{eqnarray}
Moving ZF operator $Z^{*(r)}(\beta_r)$
to the right of ZF operators
$Z^{*(r+1)}(\beta_{r+1})\cdots Z^{*(N)}(\beta_N)$,
and acting ZF operator $Z^{*(r)}(\beta_r)$
on state $|B\rangle$,
we have
\begin{eqnarray}
&&S_{r,r+1}(\beta_r-\beta_{r+1})
\cdots S_{r,N}(\beta_r-\beta_N)K_r(-\beta_r)
\nonumber\\
&\times&\langle B|{\cal Z}Z^{*(1)}(\beta_1)
\cdots Z^{*(r-1)}(\beta_{r-1})
 Z^{*(r+1)}(\beta_{r+1})\cdots Z^{*(N)}(\beta_N)
Z^{*(r)}(-\beta_r)
|B\rangle.
\end{eqnarray}
Moving ZF operator $Z^{*(r)}(-\beta_r)$
to the left and acting the dual state $\langle B|$,
we have
\begin{eqnarray}
&&
\prod_{k=1}^Q
{\cal L}(\delta_k+\beta_r)\times
S_{r,r+1}(\beta_r-\beta_{r+1})
\cdots S_{r,N}(\beta_r-\beta_N)K_r(-\beta_r)\\
&\times&
S_{N,r}(\beta_N+\beta_r)
\cdots S_{r+1,r}(\beta_{r+1}+\beta_r)
S_{r-1,r}(\beta_{r-1}+\beta_r)\cdots
S_{1,r}(\beta_1+\beta_r)K_r(-\beta_r-i\lambda)
\nonumber\\
&\times&\langle B|
Z^{*(r)}(\beta_r+2i\lambda)
{\cal Z}Z^{*(1)}(\beta_1)
\cdots Z^{*(r-1)}(\beta_{r-1})
 Z^{*(r+1)}(\beta_{r+1})\cdots Z^{*(N)}(\beta_N)
|B\rangle.\nonumber
\end{eqnarray}
Moving ZF operators $Z^{*(r)}(\beta_r)$
to the right, we have
\begin{eqnarray}
&&
\prod_{k=1}^Q
{\cal L}(\delta_k+\beta_r)\times
S_{r,r+1}(\beta_r-\beta_{r+1})
\cdots S_{r,N}(\beta_r-\beta_N)K_r(-\beta_r)\\
&\times&
S_{N,r}(\beta_N+\beta_r)
\cdots S_{r+1,r}(\beta_{r+1}+\beta_r)
S_{r-1,r}(\beta_{r-1}+\beta_r)\cdots
S_{1,r}(\beta_1+\beta_r)K_r(-\beta_r-i\lambda)
\nonumber\\
&\times&
\prod_{k=1}^Q
{\cal L}(\beta_r-\delta_k+2i\lambda)\times
S_{r,1}(\beta_r+2i\lambda-\beta_1)
\cdots S_{r,r-1}(\beta_r+2i\lambda-\beta_{r-1})
\nonumber\\
&\times&\langle B|
{\cal Z}Z^{*(1)}(\beta_1)
\cdots Z^{*(r-1)}(\beta_{r-1})
Z^{*(r)}(\beta_r+2i\lambda)
Z^{*(r+1)}(\beta_{r+1})\cdots Z^{*(N)}(\beta_N)
|B\rangle.\nonumber
\end{eqnarray}
Now we arrive at the following system of 
difference equations.
\begin{eqnarray}
&&f^{{\cal Z}}(\beta_1,\cdots,
\beta_{r-1},
\beta_r-2i\lambda,\beta_{r+1},\cdots,
\beta_N)
=\prod_{k=1}^Q
{\cal L}(\delta_k+\beta_r-2i\lambda)
{\cal L}(\beta_r-\delta_k)\nonumber\\
&\times&
S_{r,r+1}(\beta_r-\beta_{r+1}-2i\lambda)
\cdots S_{r,N}(\beta_r-\beta_N-2i\lambda)
K_r(-\beta_r+2i\lambda)\nonumber
\\
&\times&
S_{N,r}(\beta_N+\beta_r-2i\lambda)
\cdots S_{r+1,r}(\beta_{r+1}+\beta_r-2i\lambda)
\nonumber\\
&\times&
S_{r-1,r}(\beta_{r-1}+\beta_r-2i\lambda)\cdots
S_{1,r}(\beta_1+\beta_r-2i\lambda)K_r(-\beta_r+i\lambda)
\nonumber\\
&\times&
S_{r,1}(\beta_r-\beta_1)
\cdots S_{r,r-1}(\beta_r-\beta_{r-1})
f^{{\cal Z}}(\beta_1,\cdots,\beta_{r-1},\beta_r,
\beta_{r+1},\cdots,\beta_N
).\nonumber\\
\label{BQKZ1}
\end{eqnarray}
As the same argument as (\ref{BQKZ1}),
we get a similar equation.
\begin{eqnarray}
&&f^{{\cal Z}}(\beta_1,\cdots,\beta_{r-1},
\beta_r+2i\lambda,\beta_{r+1},
\cdots,\beta_N)=
\prod_{k=1}^Q
{\cal L}(\delta_k-\beta_r-2i\lambda)
{\cal L}(-\beta_r-\delta_k)\nonumber\\
&\times&
S_{r-1,r}(\beta_{r-1}-\beta_r-2i\lambda)
\cdots S_{1,r}(\beta_1-\beta_r-2i\lambda)
{K}_r(\beta_r+i\lambda)\nonumber\\
&\times&
S_{r,1}(-\beta_r-\beta_1) \cdots 
S_{r,r-1}(-\beta_{r}-\beta_{r-1})\nonumber\\
&\times&
S_{r,r+1}(-\beta_{r}-\beta_{r+1}) 
\cdots S_{r,N}(-\beta_r-\beta_N)K_r(\beta_r)\nonumber\\
&\times&
S_{N,r}(\beta_N-\beta_r) 
\cdots S_{r+1,r}(\beta_{r+1}-\beta_{r})
f^{{\cal Z}}(\beta_1,\cdots, \beta_{r-1}, \beta_r, 
\beta_{r+1}, \cdots, \beta_N).\nonumber\\
\label{BQKZ2}
\end{eqnarray}
We call this systems of difference equations
(\ref{BQKZ1}), (\ref{BQKZ2})
``boundary quantum Knizhnik-Zamolodchikov equations''
(boundary qKZ equations).
When we set parameter $\lambda=\pi$,
boundary qKZ equations
(\ref{BQKZ1}), (\ref{BQKZ2})
describe form factors for the affine $A_{n-1}^{(1)}$
Toda fields with boundary reflection.

\subsection{Integral Representations}

The purpose of this section is to give integral
representations of form factors $f^{{\cal Z}}
(\beta_1,\cdots,\beta_N)$.
In order to evaluate the above vacuum expectation value,
we invoke free field realization of 
various quantities,
the Zamolodchikov-
Faddeev operators $Z_j^*(\beta)$, the local operators
$Z_k'(\delta)$, the boundary state $|B\rangle$
and its dual $\langle B|$.

At first we present results of
simple cases.
\begin{eqnarray}
\frac{\langle B|
Z_0'(\delta)
|B\rangle}{\langle B|B \rangle}
=\exp
\left(\int_C
\frac{dt}{2\pi i t}{\rm log}(-t)
\frac{1}{1-e^{-\lambda t}}
\frac{{\rm ch}(\frac{\rho t}{4})
}{
{\rm sh}(\frac{\rho}{4}+\frac{\pi}{2n})t
}
\frac{{\rm sh}\frac{\pi t}{2}
{\rm sh}\frac{(n-1)\pi t}{2}
}{{\rm sh}\pi t}
(e^{-\lambda t-i\delta t}+e^{i\delta t})\right.
\nonumber\\
\left.
-\int_C\frac{dt}{2\pi i t}{\rm log}(-t)
\frac{1}{1-e^{-2\lambda t}}
\frac{{\rm sh}(\frac{\rho t}{2})}
{{\rm sh}(\frac{\rho}{2}+\frac{\pi}{n})t
}
\frac{
{\rm sh}\frac{(n-1)\pi t}{n}
}{{\rm sh}\pi t}
(\frac{1}{2}e^{-2\lambda t-2i\delta t}+
e^{-2\lambda t}+\frac{1}{2}e^{2i\delta t})
\right), \nonumber\\
\label{1point}
\end{eqnarray}
and
\begin{eqnarray}
&&f^{Z_0'(\delta)}(\beta)_0=\frac{\langle B|
Z_0'(\delta)Z_0^*(\beta)|B\rangle}
{\langle B|B \rangle}\nonumber\\
&=&
e^{-(\gamma+{\rm log}2\pi)\frac{n-1}{n}}
\frac{\Gamma(\frac{\delta-\beta}{2\pi i}+\frac{1}{2n})}
{\Gamma(\frac{\delta-\beta}{2\pi i}+1-\frac{1}{2n}
)}\times
\exp\left(-
\int_C
\frac{dt}{2\pi i t}{\rm log}(-t)
\frac{1}{1-e^{-\lambda t}}
\frac{{\rm sh}\frac{\pi t}{2}
{\rm sh}\frac{(n-1)\pi t}{2n}
}{{\rm sh}\pi t}
\right.
\nonumber\\
&\times&\left.
\left\{
\frac{{\rm ch}(\frac{\rho}{4}+\frac{\pi}{2n})t}
{{\rm sh}(\frac{\rho t}{4})}
(e^{i\beta t}+e^{-\lambda t-i\beta t})
-
\frac{{\rm ch}(\frac{\rho t}{4})}
{{\rm sh}(\frac{\rho}{4}+\frac{\pi}{2n})t}
(e^{i\delta t}+e^{-\lambda t-i\delta t})
\right\}\right.\nonumber\\
&+&\left.
\int_C
\frac{dt}{2\pi i t}{\rm log}(-t)
\frac{1}{1-e^{-2\lambda t}}
\frac{{\rm sh}
\frac{(n-1)\pi t}{n}}{{\rm sh}\pi t}
\left\{
(e^{-2\lambda t-i\beta t-i \delta t}
+e^{-2\lambda t+i\beta t-i \delta t}+
e^{-2\lambda t-i\beta t+i \delta t}+
e^{i\beta t+i\delta t}
)
\right.\right.\nonumber\\
&-&\left.\left.
\frac{{\rm sh}(\frac{\rho t}{2})}{
2{\rm sh}(\frac{\rho}{2}+\frac{\pi}{n})t 
}(e^{-2\lambda t-2i\delta t}+2e^{-2\lambda t}+e^{2i\delta t})
-
\frac{{\rm sh}(\frac{\rho}{2}+\frac{\pi}{n})t}{
2{\rm sh}(\frac{\rho t}{2}) 
}(e^{-2\lambda t-2i\beta t}+2e^{-2\lambda t}+e^{2i\beta t})
\right\}
\right).\nonumber
\\
\label{special}
\end{eqnarray}
Here the integrand contour $C$ is
given in Appendix A.
It is easily seen that
the above formula is re-written by multiple Gamma
functions $\Gamma_r(x|\omega_1 \cdots \omega_r)$
summarized in Appendix A.
When we set $\lambda=\pi$, we get form factor of
the affine $A_{n-1}$ Toda fields with boundary reflection.
In a limiting case $\rho \to \infty$,
our formula reproduce form factor for the
$SU(n)$ invariant massive Thirring model
with boundary reflection
\cite{K1, FKQ}.
When we set $\lambda \to 0$,
the quantity
$f^{\cal Z}(\beta_1, \cdots , \beta_N)$ 
produces an eigenvector of $A_{n-1}^{(1)}$
analogue of finite XXZ chain with double boundaries
\cite{FW}.

Next we present general formulae
of the form factors $f^{{\cal Z}}(\beta_1,\cdots,\beta_N)
_{j_1,\cdots,j_N}$, and, at the same time,
explain how to evaluate 
the vacuum expectation values.
Let us fix
the indexes $\{j_1,\cdots,j_N\}$, where 
$j_1,\cdots,j_N \in
\{0,1,\cdots,n-1\}$, and 
 $\{k_1,\cdots,k_Q\}$, where 
$k_1,\cdots,k_Q \in
\{0,1,\cdots,n-1\}$.
We associate the integration variables
$\alpha_{j,r}, (1\leq r \leq N, 1\leq j \leq j_r)$
to the basic operator $V_j(\alpha_{j,r})$
contained in the Zamolodchikov-Faddeev operator
$Z_{j_r}^*(\beta_r)$.
We also use the notation $\alpha_{0,r}=\beta_r$.
We associate the integration variables
$\gamma_{k,s}, (1\leq s \leq Q, 1\leq k \leq k_s)$
to the basic operator $U_k(\alpha_{k,s})$
contained in the local operators
$Z_{k_s}'(\delta_s)$.
We also use the notation $\gamma_{0,s}=\delta_s$. 
Let us set the index set
${\cal A}_j$ and ${\cal G}_k$ by
\begin{eqnarray}
{\cal A}_j=\{r|j_r \geq j\},~~~
{\cal G}_k=\{s|k_s \geq k\}.
\end{eqnarray}
By normal-ordering the product of the Zamolodchikov-Faddeev
operators and the local operators,
we have the following formulae.
\begin{eqnarray}
&&f^{{\cal Z}}(\beta_1,\cdots,\beta_N)_{j_1,\cdots,j_N}
\nonumber\\
&=&
E(\{\beta\}|\{\delta\})
\prod_{r=1}^N
\prod_{j=1}^{j_r}
\int_{C_j^*}
d\alpha_{j,r}
\prod_{s=1}^Q\prod_{k=1}^{k_s}
\int_{-\infty}^\infty d\gamma_{k,s}
I(\{\alpha\}|\{\gamma\})_{j_1,\cdots,j_N}^{k_1,\cdots,k_Q}.
\label{Representation}
\end{eqnarray}
Here the integral contour 
$C\j^*$ was defined below (\ref{real:Psi*}).\\
Here we set the leading factor
$E(\{\beta\}|\{\delta\})$ by
\begin{eqnarray}
&&E(\{\beta\}|\{\delta\})
=e^{-\frac{\pi}{\rho+\frac{2\pi}{n}}(
\delta_1+\cdots+\delta_Q)
+\frac{\pi}{\rho}(\beta_1+\cdots+\beta_N)
}
\nonumber\\
&\times&
\prod_{1\leq k_1<k_2 \leq Q}
C_{0,0}^{UU}(\delta_{k_1}-\delta_{k_2})
\prod_{1\leq j_1<j_2 \leq N}
C_{0,0}^{VV}(\beta_{j_1}-\beta_{j_2})
\prod_{k=1}^Q \prod_{j=1}^N
C_{0,0}^{UV}(\delta_k-\beta_j).\nonumber\\
\label{def:E}
\end{eqnarray}
Here we used abbreviations
$C_{0,0}^{UU}(\delta), C_{0,0}^{VV}(\beta)$, and
$C_{0,0}^{UV}(\delta)$,
which were introduced in Appendix B.
Here we set the integrand function by
\begin{eqnarray}
&&I(\{\alpha\}|\{\gamma\})_{j_1,\cdots,j_N}^{k_1,\cdots,k_Q}
\nonumber\\
&=&
\prod_{s=1}^Q 
e^{\frac{\pi}{\rho+\frac{2\pi}{n}}
\gamma_{k_s,s}}
\prod_{k=1}^{k_s}
\Gamma\left(
\frac{
i(\gamma_{k,s}-\gamma_{k-1,s})+\frac{\pi}{n}}
{\rho+\frac{2\pi}{n}}\right)
\Gamma\left(
\frac{i(\gamma_{k-1,s}-\gamma_{k,s})+\frac{\pi}{n}}
{\rho+\frac{2\pi}{n}}\right)\nonumber\\
&\times&
\prod_{r=1}^N 
e^{-\frac{\pi}{\rho}\alpha_{j_r,r}}
\prod_{j=1}^{j_r}
\Gamma\left(
\frac{
i(\alpha_{j,r}-\alpha_{j-1,r})-\frac{\pi}{n}}
{\rho}\right)
\Gamma\left(
\frac{i(\alpha_{j-1,r}-\alpha_{j,r})-\frac{\pi}{n}}
{\rho}\right)\nonumber\\
&\times&\prod_{k=1}^{n-1}
\prod_{s_1 \in {\cal G}_k,s_2 \in {\cal G}_{k-1}
\atop{s_1<s_2}}
C_{k,k-1}^{UU}(\gamma_{k,s_1}-
\gamma_{k-1,s_2})
\prod_{s_1 \in {\cal G}_{k-1},s_2 \in {\cal G}_{k}
\atop{s_1<s_2}}
C_{k-1,k}^{UU}(\gamma_{k-1,s_1}-
\gamma_{k,s_2})\nonumber\\
&\times&
\prod_{k=1}^{n-1}
\prod_{s_1,s_2 \in {\cal G}_k
\atop{s_1<s_2}}
C_{k,k}^{UU}(\gamma_{k,s_1}-
\gamma_{k,s_2})
\prod_{j=1}^{n-1}
\prod_{r_1,r_2 \in {\cal A}_j
\atop{r_1<r_2}}
C_{j,j}^{VV}(\alpha_{j,r_1}-
\alpha_{j,r_2})\nonumber\\
&\times&
\prod_{j=1}^{n-1}
\prod_{r_1 \in {\cal A}_j,r_2 \in {\cal A}_{j-1}
\atop{r_1<r_2}}
C_{j,j-1}^{VV}(\alpha_{j,r_1}-
\alpha_{j-1,r_2})
\prod_{r_1 \in {\cal A}_{j-1},r_2 \in {\cal A}_{j}
\atop{r_1<r_2}}
C_{j-1,j}^{VV}(\alpha_{j-1,r_1}-
\alpha_{j,r_2})\nonumber\\
&\times&
\prod_{k=1}^{n-1}
\prod_{s\in{\cal G}_k,r\in{\cal A}_k}
C_{k,k}^{UV}(\gamma_{k,s}-\alpha_{k,r})
\prod_{
s\in{\cal G}_k,r\in{\cal A}_{k-1}
}
C_{k,k-1}^{UV}(\gamma_{k,s}-\alpha_{k-1,r})\nonumber\\
&\times&\prod_{k=1}^{n-1}
\prod_{
s\in{\cal G}_{k-1},r\in{\cal A}_{k}
}
C_{k-1,k}^{UV}(\gamma_{k-1,s}-\alpha_{k,r})
\times
J(\{\alpha\}|\{\gamma\})_{j_1,\cdots,j_N}^{k_1,\cdots,k_Q}.
\label{def:I}
\end{eqnarray}
Here we used abbereviations $C_{j_1,j_2}^{VV}(\beta),
C_{j_1,j_2}^{UU}(\beta)$ and $C_{j_1,j_2}^{UV}(\beta)$,
which were
introduced in Appendix B.
Here we have set
\begin{eqnarray}
&&J(\{\alpha\}|\{\gamma\})_{j_1,\cdots,j_N}^{k_1,\cdots,k_Q}
\\
&=&
\frac{1}{\langle B|B \rangle}\times
\langle B| 
\exp\left(
\int_0^\infty
\sum_{j=1}^{n-1}X_j(t)a_j(-t)dt
\right)
\exp\left(
\int_0^\infty
\sum_{j=1}^{n-1}Y_j(t)a_j(t)dt
\right)|B\rangle,\nonumber
\end{eqnarray}
where
\begin{eqnarray}
X_j(t)=
\frac{{\rm sh}\frac{(n-j)\pi t}{n}}{{\rm sh}\pi t}
\left(-\sum_{r=1}^Ne^{-i\beta_r t}+
\frac{{\rm sh}(\frac{\rho t}{2})}
{{\rm sh}(\frac{\rho}{2}+\frac{\pi}{n})t}
\sum_{s=1}^Q e^{-i\delta_s t}\right)\nonumber\\
+\sum_{r \in {\cal A}_j}e^{-i\alpha_{j,r}t}-
\frac{{\rm sh}(\frac{\rho t}{2})}
{{\rm sh}(\frac{\rho}{2}+\frac{\pi}{n})t}
\sum_{s \in {\cal G}_j}
e^{-i\gamma_{j,s}t},\label{X}
\\
Y_j(t)=
\frac{{\rm sh}\frac{(n-j)\pi t}{n}}{{\rm sh}\pi t}
\left(-\sum_{r=1}^Ne^{i\beta_r t}+
\frac{{\rm sh}(\frac{\rho t}{2})}
{{\rm sh}(\frac{\rho}{2}+\frac{\pi}{n})t}
\sum_{s=1}^Q e^{i\delta_s t}\right)\nonumber\\
+\sum_{r \in {\cal A}_j}e^{i\alpha_{j,r}t}-
\frac{{\rm sh}(\frac{\rho t}{2})}
{{\rm sh}(\frac{\rho}{2}+\frac{\pi}{n})t}
\sum_{s \in {\cal G}_j}
e^{i\gamma_{j,s}t}.\label{Y}
\end{eqnarray}

Next we evaluate the vacuum expectation value,
$J(\{\alpha\}|\{\gamma\})_{j_1,\cdots,j_N}
^{k_1,\cdots,k_Q}$, and get
a formulae without free field operators.
For our purpose
we use the coherent state, $|\xi_1,\cdots,\xi_{n-1}\rangle$
and its dual state $\langle \bar{\xi}_1,
\cdots, \bar{\xi}_{n-1}|$, defined by
\begin{eqnarray}
|\xi_1,\cdots,\xi_{n-1}\rangle=
\exp\left(\sum_{k=1}^{n-1}
\int_0^\infty \xi_k(s)a_k(-s)ds\right)
|vac\rangle,\\
\langle \bar{\xi}_1,\cdots,
\bar{\xi}_{n-1}|=\langle vac|
\exp\left(
\sum_{k=1}^{n-1}
\int_0^\infty
\bar{\xi}_k(s)a_k(s)ds
\right).
\end{eqnarray}
The coherent state enjoy
\begin{eqnarray}
a_j(t)|\xi_1,\cdots,\xi_{n-1}\rangle&=&
\sum_{k=1}^{n-1}A_{j,k}(t)\xi_{k}(t)|
\xi_1,\cdots,\xi_{n-1}\rangle,~~(t>0),
\\
\langle \bar{\xi}_1,\cdots,\bar{\xi}_{n-1}|
a_j(-t)&=&
\sum_{k=1}^{n-1}A_{j,k}(t)\bar{\xi}_k(t)\langle
\bar{\xi}_1,\cdots,\bar{\xi}_{n-1}|,~~(t>0).
\end{eqnarray}
The following completeness relation by means of
Feynmann path integral is useful.
\begin{eqnarray}
id&=&c_F \times \int \prod_{k=1}^{n-1}
\prod_{s>0}d\xi_k(s)d\bar{\xi}_k(s)\nonumber\\
&\times&
\exp\left(
-\sum_{k_1,k_2=1}^{n-1}
\int_0^\infty
A_{k_1,k_2}(s)\xi_{k_1}(s)\bar{\xi}_{k_2}(s)ds\right)
|\xi_1,\cdots,\xi_{n-1}\rangle
\langle \bar{\xi}_1,\cdots,\bar{\xi}_{n-1}|.
\end{eqnarray}
Here the integration $\int d\xi d\bar{\xi}$
is taken over the entire complex plane
with the measure
$d\xi d\bar{\xi}=-2i dxdy$
for $\xi=x+iy$.
Here $c_F$ is a constant.
\\
In what follows we use the abberiviations,
$\tilde{\beta}_j(t), \tilde{\delta}_j(t)$,
$\widetilde{X}_j(t)$, and $\widetilde{Y}_j(t)$
defined by
\begin{eqnarray}
\tilde{\beta}_j(t)&=&
{\rm sh}
\left(\frac{\pi t}{2n}\right)\times
\frac{
\displaystyle
{\rm ch}
\left(
\left(\frac{\rho}{4}+\frac{\pi}{2n}\right)t\right)}
{\displaystyle
{\rm sh}\left(\frac{\rho t}{4}\right)
}=t\times
\sum_{k=1}^{n-1}A_{j,k}(t)\beta_k(t),
\\
\tilde{\delta}_j(t)&=&
e^{-\lambda t}\times
\tilde{\beta}_j(t)=t\times
\sum_{k=1}^{n-1}A_{j,k}(t)\delta_k(t),
\end{eqnarray}
\begin{eqnarray}
&&\widetilde{X}_j(t)=t\times
\sum_{k=1}^{n-1}A_{j,k}(t)X_k(t)
\label{tX}
\\
&=&
\frac{{\rm sh}(\frac{\rho}{2}+\frac{\pi}{n})t}
{{\rm sh}(\frac{\rho t}{2})}
\times
\left\{
\sum_{r \in {\cal A}_{j-1}}
e^{-i\alpha_{j-1,r}t}
+
\sum_{r \in {\cal A}_{j+1}}
e^{-i\alpha_{j+1,r}t}
-(e^{\frac{\pi}{n}t}+e^{-\frac{\pi}{n}t}
)\sum_{r \in {\cal A}_j}
e^{-i\alpha_{j,r}t}
\right\}\nonumber\\
&-&
\left\{
\sum_{s \in {\cal G}_{j-1}}
e^{-i\gamma_{j-1,s}t}
+
\sum_{s \in {\cal G}_{j+1}}
e^{-i\gamma_{j+1,s}t}
-(e^{\frac{\pi}{n}t}+e^{-\frac{\pi}{n}t}
)\sum_{s \in {\cal G}_j}
e^{-i\gamma_{j,s}t}
\right\},
\nonumber
\end{eqnarray}
and
\begin{eqnarray}
&&\widetilde{Y}_j(t)=
t\times
\sum_{k=1}^{n-1}A_{j,k}(t)Y_k(t)
\label{tY}
\\
&=&
\frac{{\rm sh}(\frac{\rho}{2}+\frac{\pi}{n})t}
{{\rm sh}(\frac{\rho t}{2})}
\times
\left\{
\sum_{r \in {\cal A}_{j-1}}
e^{i\alpha_{j-1,r}t}
+
\sum_{r \in {\cal A}_{j+1}}
e^{i\alpha_{j+1,r}t}
-(e^{\frac{\pi}{n}t}+e^{-\frac{\pi}{n}t}
)\sum_{r \in {\cal A}_j}
e^{i\alpha_{j,r}t}
\right\}\nonumber\\
&-&
\left\{
\sum_{s \in {\cal G}_{j-1}}
e^{i\gamma_{j-1,s}t}
+
\sum_{s \in {\cal G}_{j+1}}
e^{i\gamma_{j+1,s}t}
-(e^{\frac{\pi}{n}t}+e^{-\frac{\pi}{n}t}
)\sum_{s \in {\cal G}_j}
e^{
i\gamma_{j,s}t}
\right\}.
\nonumber
\end{eqnarray}
Here we understand ${\cal A}_n,
{\cal G}_n=\phi$.\\
Changing the order of the bosonic operators,
we have
\begin{eqnarray}
&&J(\{\alpha\}|\{\gamma\})_{j_1,\cdots,j_N}
^{k_1,\cdots,k_Q}\nonumber\\
&=&
\frac{1}{\langle vac |e^{G}e^{B}|vac \rangle}\times
\langle vac |e^G
\exp\left(
\int_0^\infty
\sum_{j=1}^{n-1}Y_j(t)a_j(t)dt
\right)
\exp\left(
\int_0^\infty
\sum_{j=1}^{n-1}X_j(t)a_j(-t)dt
\right)
e^B
|vac\rangle \nonumber\\
&\times&
\exp\left(
\int_C
\frac{{\rm log}(-t)}{2\pi i t}
\frac{\displaystyle{\rm sh}\left(\frac{\rho t}{2}\right)}
{\displaystyle{\rm sh}\left(\left(\frac{\rho}{2}+\frac{\pi}{n}\right)t\right)}
\sum_{l_1,l_2=1}^{n-1}\tilde{X}_{l_1}(t)
I_{l_1,l_2}(t)
\tilde{Y}_{k_l}(t)
dt\right).
\end{eqnarray}
When we insert the completeness relation of the coherent
states, we have without-operator 
formula.
\begin{eqnarray}
&&\langle vac |
e^G  
\exp\left(
\int_0^\infty
\sum_{j=1}^{n-1}Y_j(t)a_j(t)dt
\right)
\exp\left(
\int_0^\infty
\sum_{j=1}^{n-1}X_j(t)a_j(-t)dt
\right)
|vac \rangle
\nonumber\\
&=& c_F \times
\int \prod_{j=1}^{n-1}\prod_{t>0}
d\xi_j(t)d\bar{\xi}_j(t)
\nonumber\\
&\times&
\exp\left(
\int_0^\infty
\frac{1}{t}
\sum_{j=1}^{n-1}
(\widetilde{Y}_j(t)+\tilde{\delta}_j(t))\xi_j(t)dt+
\int_0^\infty
\frac{1}{t}
\sum_{j=1}^{n-1}
(\widetilde{X}_j(t)+\tilde{\beta}_j(t))
\bar{\xi}_j(t)dt
\right.
\nonumber\\
&+&\left.
\int_0^\infty
\sum_{l_1,l_2=1}^{n-1}
A_{l_1,l_2}(t)
\left(\frac{e^{-2\lambda t}}{2}
\xi_{l_1}(t)\xi_{l_2}(t)-\xi_{l_1}(t)\bar{\xi}_{l_2}(t)
+\frac{1}{2}\bar{\xi}_{l_1}(t)\bar{\xi}_{l_2}(t)
\right)dt
\right).\nonumber\\
\label{J'}
\end{eqnarray}

Performing the Gauss integral calculation,
$\int_{-\infty}^\infty e^{-x^2}dx=\sqrt{\pi}$,
(\ref{J'}) becomes the following, up to constant multiplicity.
\begin{eqnarray}
&&\exp\left(
-\int_C \frac{dt}{2\pi i t}{\rm log}(-t)
\frac{1}{1-e^{-2\lambda t}}
\frac{{\rm sh}\left(\frac{\rho t}{2}\right)}
{{\rm sh}\left(\frac{\rho}{2}+\frac{\pi}{n}\right)t}\right.
\nonumber\\
&\times&
\left(\sum_{l_1,l_2=1}^{n-1}
I_{l_1,l_2}(t)\left(
\frac{e^{-2\lambda t}}{2}
(\widetilde{X}_{l_1}(t)+
\widetilde{\beta}_{l_1}(t))
(\widetilde{X}_{l_2}(t)+
\widetilde{\beta}_{l_2}(t))
\right.\right.
\nonumber\\
&+&\left.
\left.\left.
(\widetilde{X}_{l_1}(t)+
\widetilde{\beta}_{l_1}(t))
(\widetilde{Y}_{l_2}(t)+
\widetilde{\delta}_{l_2}(t)
)
+\frac{1}{2}
(\widetilde{Y}_{l_1}(t)+
\widetilde{\delta}_{l_1}(t))
(\widetilde{Y}_{l_2}(t)+
\widetilde{\delta}_{l_2}(t))
\right)\right)
\right).\nonumber
\\
\end{eqnarray}

Now we arrive at the following formula.
\begin{eqnarray}
&&J(\{
\alpha\}|\{\gamma\})
_{j_1,\cdots,j_N}^{k_1,\cdots,k_Q}\nonumber\\
&=&
\exp\left(-
\int_{C}
\frac{dt}{2\pi i t}{\rm log}(-t)
\frac{1}{1-e^{-2 \lambda t}}
\frac{\displaystyle {\rm sh}\left(\frac{\rho t}{2}\right)}{
\displaystyle {\rm sh}\left(\left(
\frac{\rho}{2}+\frac{\pi}{n}\right)t\right)
}
\frac{1}{\displaystyle {\rm sh}\left(\frac{\pi t}{n}\right)
\displaystyle
{\rm sh}\left(\pi t\right)
}
\right.\nonumber\\
&\times&\left.
\left(
\sum_{l=1}^{n-1}
{\rm sh}\left(\frac{\pi l t}{n}\right)
{\rm sh}\left(\frac{(n-l)\pi t}{n}\right)
\left(
\frac{e^{-2\lambda t}}{2}\widetilde{X}_l(t)^2+
e^{-2\lambda t}\widetilde{X}_l(t)
\widetilde{Y}_l(t)+\frac{1}{2}
\widetilde{Y}_l(t)^2
\right)
\right.\right.\nonumber\\
&+&\left.
\left.
\sum_{1\leq l_1<l_2 \leq n-1}
{\rm sh}\left(\frac{\pi l_1 t}{n}\right)
{\rm sh}\left(\frac{(n-l_2)\pi t}{n}\right)
(
e^{-2\lambda t}
\widetilde{X}_{l_1}(t)
\widetilde{X}_{l_2}(t)+
e^{-2\lambda t}\widetilde{X}_{l_1}(t)
\widetilde{Y}_{l_2}(t)
\right.\right.\nonumber\\
&+&\left.\left.
e^{-2\lambda t}\widetilde{X}_{l_2}(t)
\widetilde{Y}_{l_1}(t)
+
\widetilde{Y}_{l_1}(t)
\widetilde{Y}_{l_2}(t)
)
\right)
\right.-\left.
\int_{C}
\frac{dt}{2\pi i t}{\rm log}(-t)
\frac{1}{1-e^{-\lambda t}}
\frac{\displaystyle
{\rm ch}\left(\frac{\rho t}{4}\right)
}{\displaystyle
{\rm sh}\left(\left(\frac{\rho}{4}+\frac{\pi}{2n}\right)t\right)
}
\right.\nonumber\\
&\times&\left.
\frac{
\displaystyle
{\rm sh}\left(\frac{\pi t}{2}\right)}{
\displaystyle
{\rm sh}\left(\frac{\pi t}{2n}\right)
{\rm sh}(\pi t)
}
\sum_{l=1}^{n-1}
{\rm sh}\left(\frac{l\pi t}{2n}\right)
{\rm sh}\left(\frac{(n-l)\pi t}{2n}\right)
(e^{-\lambda t}
\widetilde{X}_l(t)+\widetilde{Y}_l(t))
\right).\nonumber\\
\label{result:J}
\end{eqnarray}
Here $\widetilde{X}_l(t),
\widetilde{Y}_l(t)$ are defined in
(\ref{tX}), (\ref{tY}).
Here the integral contour $C$ is as the same as
those given in Appendix A.
By this result,
it is easily seen that
$J(\{\alpha\}|\{\gamma\})_{j_1,\cdots.j_N}
^{k_1,\cdots,k_Q}$
is evaluated by multi-Gamma functions
$\Gamma_r(x|\omega_1 \cdots \omega_r)
$,
summarized in Appendix A.
In a limiting case $\rho \to \infty$,
our integral formulae reproduce those of form factors
for the $SU(n)$
invariant massive Thirring model with boundary reflection
\cite{K1}.

~\\
{\it Let us summarize the result of this section.
We present integral representations
(\ref{Representation}) for form factors of the local fields.
}
\begin{eqnarray}
&&f^{{\cal Z}}
(\beta_1,\cdots,\beta_N)_{j_1,\cdots,j_N}\nonumber\\
&=&
E(\{\beta\}|\{\delta\})
\prod_{r=1}^N\prod_{j=1}^{j_r}
\int_{C_j^*}
d\alpha_{j,r}
\prod_{s=1}^Q\prod_{k=1}^{k_s}
\int_{-\infty}^\infty
d\gamma_{k,s}
I(\{\alpha\}|\{\gamma\})
_{j_1,\cdots,j_N}^{k_1,\cdots,k_Q}.\label{main}
\end{eqnarray}
{\it Here the factor $E(\{\beta\}|\{\delta\})$
is given by (\ref{def:E}).
The integrand $I(\{\alpha\}|\{\gamma\})
_{j_1,\cdots,j_N}^{k_1,\cdots,k_Q}$
is given in (\ref{def:I}),
where the factor $J(\{\alpha\}|\{\gamma\})
_{j_1,\cdots,j_N}^{k_1,\cdots,k_Q}$ is given in (\ref{result:J}),
and $C_{j_1,j_2}^{UU}(\alpha), C_{j_1,j_2}^{VV}(\alpha)$
and $C_{j_1,j_2}^{UV}(\alpha)$
are given in
Appendix B.
The integral contour $C_j^*$ is given below 
(\ref{real:Psi*}).
For special cases
we present more explicit formulae in 
(\ref{1point}),
(\ref{special}).
}

~\\
Fateev et al.\cite{FLZZ} proposed
an expression for vacuum expectation values of the
special field,
for boundary SG model,
which is described by the following action
\begin{eqnarray}
{\cal A}_{FLZZ}=
\int_{-\infty}^\infty
dx\int_{0}^\infty
dy
\left((\partial_x \varphi)^2+(\partial_y \varphi)^2
\right)-\mu \int_{-\infty}^\infty
dx~{\rm cos}(\beta \varphi(x,0)).
\end{eqnarray}
In this paper
and Hou et al.\cite{HSWY},
boundary SG model
(affine $A_1^{(1)}$ Toda fields)
is the model
which is described by the following action
\begin{eqnarray}
{\cal A}_{BSG}=
\int_{-\infty}^\infty
dx\int_{0}^\infty
dy
\left((\partial_x \varphi)^2+(\partial_y \varphi)^2
-\mu~{\rm cos}(\beta \varphi)\right).
\end{eqnarray}
Our considering boundary SG model 
${\cal A}_{BSG}$
is different from
Fateev et al.'s model ${\cal A}_{FLZZ}$.
V.Fateev, A.Zamolodchikov, Al.Zamolodchikov
\cite{FZZ} studied Boundary Liouville
conformal field theory.
For a particular application,
they present one point function of
the special operator in the boundary SG
model ${\cal A}_{BSG}$.
To reveal the connection between two formulae
\cite{HSWY} and \cite{FZZ} is 
our future problem.

~\\
{\bf Acknowledgement}
~~This work was partly supported by Grant-in-Aid
for Encouragements for Young Scientists ({\bf A})
from JSPS (11740099).

\begin{appendix}

\section{Multiple Gamma function}

Here we summarize the multiple gamma and the multiple sine
functions, following \cite{JM}.
Let us set the multiple Gamma function
$\Gamma_r(x|\omega_1 \cdots \omega_r)$
by
\begin{eqnarray}
{\rm log}\Gamma_r(x|\omega_1 \cdots \omega_r)
=
\frac{(-1)^r}{r!}\gamma
B_{r,r}(x|\omega_1 \cdots \omega_r)
+\int_C
\frac{e^{-xt}{\rm log}(-t)}{
\prod_{j=1}^r
(1-e^{-\omega_j t})}
\frac{dt}{2\pi i t},
\end{eqnarray}
where
the functions $B_{jj}(x)$ are the multiple Bernoulli polynomials
defined by
\begin{eqnarray}
\frac{t^r e^{xt}}{
\prod_{j=1}^r (e^{\omega_j t}-1)}=
\sum_{n=0}^\infty
\frac{t^n}{n!}B_{r,n}(x|\omega_1 \cdots \omega_r).
\end{eqnarray}
Here $\gamma$ is Euler's constant,
$\gamma=\lim_{n\to \infty}
(1+\frac{1}{2}+\frac{1}{3}+\cdots+\frac{1}{n}-{\rm log}n)$.\\
Here the contor of integral is given by

~\\
~\\

\unitlength 0.1in
\begin{picture}(34.10,11.35)(17.90,-19.35)
%
\special{pn 8}%
\special{pa 5200 800}%
\special{pa 2190 800}%
\special{fp}%
\special{sh 1}%
\special{pa 2190 800}%
\special{pa 2257 820}%
\special{pa 2243 800}%
\special{pa 2257 780}%
\special{pa 2190 800}%
\special{fp}%
\special{pa 2190 1600}%
\special{pa 5190 1600}%
\special{fp}%
\special{sh 1}%
\special{pa 5190 1600}%
\special{pa 5123 1580}%
\special{pa 5137 1600}%
\special{pa 5123 1620}%
\special{pa 5190 1600}%
\special{fp}%
%
\special{pn 8}%
\special{pa 5190 1200}%
\special{pa 2590 1210}%
\special{fp}%
\put(25.9000,-12.1000){\makebox(0,0)[lb]{$0$}}%
%
\special{pn 8}%
\special{ar 2190 1210 400 400  1.5707963 4.7123890}%
\put(33.9000,-20.2000){\makebox(0,0){{\bf Contour} $C$}}%
\end{picture}%

~\\

Let us set the multiple sine function 
$S_r(x|\omega_1 \cdots \omega_r)$ by
\begin{eqnarray}
S_r(x|\omega_1 \cdots \omega_r)&=&
\Gamma_r(x|\omega_1 \cdots \omega_r)^{-1}
\Gamma_r(\omega_1+\cdots+\omega_r-x|
\omega_1 \cdots \omega_r)^{(-1)^r}.
\end{eqnarray}
The multiple Gamma function and
the multiple sine function satisfy
the recursion relations,
\begin{eqnarray}
\frac{
\Gamma_{r}(x+\omega_1|\omega_1 \cdots \omega_r)
}{
\Gamma_{r}(x|\omega_1 \cdots \omega_r)
}&=&\frac{1}{\Gamma_{r-1}(x|\omega_2 \cdots \omega_r)},\\
\frac{
S_{r}(x+\omega_1|\omega_1 \cdots \omega_r)
}{
S_{r}(x|\omega_1 \cdots \omega_r)
}&=&\frac{1}{S_{r-1}(x|\omega_2 \cdots \omega_r)}.
\end{eqnarray}
Here we understand $\Gamma_0=x$.
We have the following conditions.
\begin{eqnarray}
\Gamma_1(x|\omega)=
\omega^{\frac{x}{\omega}-\frac{1}{2}}
\frac{\Gamma(\frac{x}{\omega})}{\sqrt{2\pi}},
~
S_1(x|\omega)=2 {\rm sin}(\frac{\pi x}{\omega}).
\end{eqnarray}
We have
\begin{eqnarray}
\lim_{\rho \to \infty}S_1(x|\rho)=e^\gamma x,~
\lim_{\rho \to \infty}S_2(x|\rho,\omega)=\frac{
(2\pi)^{\frac{1}{2}}(\omega e^\gamma)^{\frac{1}{2}-\frac{x}{\omega}}}{\Gamma\left(\frac{x}{\omega}\right)}.
\end{eqnarray}
Explicitly the multiple Bernoulli polynomials are written by
\begin{eqnarray}
&&B_{11}(x|\omega)=\frac{x}{\omega}-\frac{1}{2},\\
&&B_{22}(x|\omega_1,\omega_2)=\frac{x^2}{\omega_1 \omega_2}
-\left(\frac{1}{\omega_1}+\frac{1}{\omega_2}\right)x
+\frac{1}{2}+\frac{1}{6}\left(\frac{\omega_1}{\omega_2}
+\frac{\omega_2}{\omega_1}\right).
\end{eqnarray}

\section{Normal Ordering}

Here we list the formulas of the form
\begin{eqnarray}
X(\beta_1)Y(\beta_2)=C^{XY}(\beta_1-\beta_2)
:X(\beta_1)X(\beta_2):,
\end{eqnarray}
where $X,Y=U_j$, and $C_{XY}(\beta)$ is a meromorphic
function on ${\mathbb{C}}$.
These formulae follow from the commutation relation
of the free bosons.
When we compute the contraction of the basic
operators,
we often encounter an integral
\begin{eqnarray}
\int_0^\infty
F(t)dt,
\end{eqnarray}
which is divergent at $t=0$.
Here we adopt the following prescription
for regularization :
it should be understood as the countour integral,
\begin{eqnarray}
\int_C F(t)\frac{{\rm log}(-t)}{2\pi i}dt. 
\end{eqnarray}
Here we used the same contour $C$ as
the same as those in Appendix A.

The basic operators $U_j(\beta), V_j(\beta)$
have the contraction relations,
for $0\leq j_1,j_2 \leq n$.
\begin{eqnarray}
U_{j_1}(\beta_1)U_{j_2}(\beta_2)&=&
C_{j_1, j_2}^{U U}(\beta_1-\beta_2)
:U_{j_1}(\beta_1)U_{j_2}(\beta_2):,\\
V_{j_1}(\beta_1)V_{j_2}(\beta_2)&=&
C_{j_1, j_2}^{V V}(\beta_1-\beta_2)
:V_{j_1}(\beta_1)V_{j_2}(\beta_2):,\\
U_{j_1}(\beta_1)V_{j_2}(\beta_2)&=&
C_{j_1, j_2}^{U V}(\beta_1-\beta_2)
:U_{j_1}(\beta_1)V_{j_2}(\beta_2):,\\
V_{j_1}(\beta_1)U_{j_2}(\beta_2)&=&
C_{j_1, j_2}^{V U}(\beta_1-\beta_2)
:V_{j_1}(\beta_1)U_{j_2}(\beta_2):.
\end{eqnarray}
Here nonzero entries are given by
\begin{eqnarray}
C_{j,j}^{UU}(\alpha)=
\frac{-i\alpha}{\rho+\frac{2\pi}{n}}
e^{\frac{2\pi}{n\rho+2\pi}(\gamma+{\rm log}(\rho+\frac{2\pi}
{n}))}\times
\frac{\Gamma\left(
\frac{-i\alpha+\rho}{\rho+\frac{2\pi}{n}}\right)}
{\Gamma\left(
\frac{-i\alpha+\frac{2\pi}{n}}{\rho+\frac{2\pi}{n}}\right)}
,(j=1,\cdots,n-1),
\nonumber
\\
\\
C_{j,j-1}^{UU}(\alpha)=
C_{j-1,j}^{UU}(\alpha)=
e^{\frac{-n\rho}{n\rho+2\pi}(\gamma+{\rm log}(\rho+\frac{2\pi}{n}))}
\frac{\Gamma
\left(\frac{-i\alpha+\frac{\pi}{n}}{\rho+\frac{2\pi}{n}}
\right)}{\Gamma
\left(
\frac{-i\alpha+\rho+\frac{\pi}{n}}{
\rho+\frac{2\pi}{n}}\right)},~(j=1,\cdots,n),
\nonumber\\
\\
C_{0,0}^{UU}(\alpha)=
C_{n,n}^{UU}(\alpha)=
e^{\gamma \frac{n-1}{n}
\frac{n\rho}{n\rho+2\pi}}
\frac{\Gamma_2(-i\alpha+\rho+\frac{2\pi}{n}|
\rho+\frac{2\pi}{n},2\pi)
\Gamma_2(-i\alpha+2\pi|
\rho+\frac{2\pi}{n},2\pi)
}{
\Gamma_2(-i\alpha+\frac{2\pi}{n}|
\rho+\frac{2\pi}{n},2\pi)
\Gamma_2(-i\alpha+2\pi+\rho|
\rho+\frac{2\pi}{n},2\pi)
},\nonumber\\
\\
C_{0,n}^{UU}(\alpha)=
C_{n,0}^{UU}(\alpha)=
e^{\gamma \frac{\rho}{n\rho+2\pi}}
\frac{\Gamma_2(-i\alpha+\pi+\rho|\rho+
\frac{2\pi}{n},2\pi)
\Gamma_2(-i\alpha+\pi+\frac{2\pi}{n}|
\rho+
\frac{2\pi}{n},2\pi)
}{
\Gamma_2(-i\alpha+\pi|\rho+
\frac{2\pi}{n},2\pi)
\Gamma_2(-i\alpha+\pi+\frac{2\pi}{n}+\rho|
\rho+
\frac{2\pi}{n},2\pi)
}.\nonumber\\
\end{eqnarray}

\begin{eqnarray}
C_{j,j}^{VV}(\alpha)=
-i\rho e^{2\gamma}
e^{\frac{4\pi}{n\rho}(\gamma+{\rm log}\rho)}
\alpha \times \frac{
\Gamma\left(-\frac{i\alpha}{\rho}+1+
\frac{2\pi}{n\rho}\right)}{
\Gamma\left(-\frac{i\alpha}{\rho}-\frac{2\pi}{n\rho}\right)
}
,(j=1,\cdots,n-1),\nonumber\\
\\
C_{j,j-1}^{VV}(\alpha)=
C_{j-1,j}^{VV}(\alpha)=
e^{-\frac{2\pi+n\rho}{n\rho}(\gamma+{\rm log}\rho)}
\frac{\Gamma\left(-\frac{i\alpha}{\rho}-
\frac{\pi}{n\rho}\right)}{\Gamma
\left(-\frac{i\alpha}{\rho}+1+\frac{\pi}{n\rho}\right)
}
,(j=1,\cdots,n),\nonumber\\
\\
C_{0, 0}^{VV}(\alpha)=C_{n, n}^{VV}(\alpha)=
e^{\gamma \frac{n-1}{n}(1+\frac{2\pi}{n\rho})}
\frac{\Gamma_2(-i\alpha+\rho+\frac{2\pi}{n}|\rho,2\pi)
\Gamma_2(-i\alpha+2\pi-\frac{2\pi}{n}|\rho,2\pi)
}{
\Gamma_2(-i\alpha|\rho,2\pi)
\Gamma_2(-i\alpha+2\pi+\rho|\rho,2\pi)
},\nonumber\\
\\
C_{0, n}^{VV}(\alpha)=C_{n, 0}^{VV}(\alpha)=
e^{\gamma \frac{1}{n}(1+\frac{2\pi}{n\rho})}
\frac{\Gamma_2(-i\alpha+\pi|\rho,2\pi)
\Gamma_2(-i\alpha+\pi+\rho|\rho,2\pi)
}{
\Gamma_2(-i\alpha+\pi-\frac{2\pi}{n}|\rho,2\pi)
\Gamma_2(-i\alpha+\pi+\frac{2\pi}{n}+\rho|\rho,2\pi)
}.\nonumber\\
\end{eqnarray}

\begin{eqnarray}
C_{j,j}^{UV}(\alpha)=
-e^{-2\gamma}\left(\alpha+\frac{\pi i}{n}\right)^{-1}
\left(\alpha-\frac{\pi i}{n}\right)^{-1}
,(j=1,\cdots,n-1),\\
C_{j,j-1}^{UV}(\alpha)=
C_{j-1,j}^{UV}(\alpha)=
-i \alpha e^{\gamma},~~~
 (j=1,\cdots,n),
\\
C_{0,0}^{UV}(\alpha)=
C_{n,n}^{UV}(\alpha)=
e^{-(\gamma+{\rm log}2\pi)\frac{n-1}{n}}
\frac{\Gamma\left(
\frac{\alpha}{2\pi i}+\frac{1}{2n}
\right)}{\Gamma\left(
\frac{\alpha}{2\pi i}+1-\frac{1}{2n}\right)},\\
C_{0,n}^{UV}(\alpha)=
C_{n,0}^{UV}(\alpha)=e^{-\frac{1}{n}(\gamma+{\rm log}2\pi)}
\frac{\Gamma\left(
\frac{\alpha}{2\pi i}+\frac{1}{2}-\frac{1}{2n}\right)}{
\Gamma\left(\frac{\alpha}{2\pi i}+\frac{1}{2}
+\frac{1}{2n}\right)},
\end{eqnarray}
and
\begin{eqnarray}
C_{j_1, j_2}^{VU}(\beta)=C^{UV}_{j_1, j_2}(\beta),
~(0\leq j_1,j_2 \leq n).
\end{eqnarray}
Here we have set the supplemental basic operators
$U_{n}(\alpha)$ by
\begin{eqnarray}
U_n(\alpha)=:\exp\left(-\int_0^\infty
b_{n-1}^*(t)e^{i\beta t}dt\right):,
~
b_{n-1}^*(t)=-\sum_{j=1}^{n-1}b_j(t)
\frac{{\rm sh}\frac{j\pi t}{n}}{
{\rm sh}\pi t}.
\end{eqnarray}

\section{$U_q(\widehat{sl_2})$ Case}
In this Appendix we give an additional
result for $U_q(\widehat{sl_2})$ case,
and give some comments for $U_q(\widehat{sl_n})$
case. 
Let us introduce the operators 
$\widehat{\Psi}_j^*(\beta), (j=0,1)$ by
\begin{eqnarray}
\widehat{\Psi}_0^*(\beta)&=&
V_0(\beta),\\
\widehat{\Psi}_1^*(\beta)&=&
\int_{C_1^*} d\alpha :V_0(\beta)V_1(\alpha):
e^{\frac{\pi}{\rho}\beta}~
\Gamma\left(
\frac{i(\beta-\alpha)}{\rho}-\frac{\pi}{2\rho}\right)
\Gamma\left(
\frac{i(\alpha-\beta)}{\rho}-\frac{\pi}{2\rho}
\right),
\end{eqnarray}
where 
the integral contour $C_1^*$ is given in section 3.
The operators 
$\widehat{\Psi}_j^*(\beta)$ is slightly
different from
realizations of $Z_j^*(\beta)$
(\ref{real:Psi*}).
Factor $e^{-\frac{\alpha}{\rho}}$
drops in $\widehat{\Psi}_j^*(\beta)$.
However the operators
$\widehat{\Psi}_j^*(\beta)$ 
satisfy the same commutation relation
(\ref{Com1}), too.
Let us introduce the boundary $K$-matrix
$\widehat{K}(\beta) \in {\rm End}(
{\mathbb{C}}^2)$ by
\begin{eqnarray}
\widehat{K}(\beta)=
\frac{\hat{G}_0(-\beta)}{\hat{G}_0(\beta)
}\left(
\begin{array}{cc}
1&0\\
0&\frac{e^{-\frac{2\pi}{\rho}\beta}-
e^{\frac{2\pi}{\rho}\mu}}{
e^{\frac{2\pi}{\rho}\beta}-
e^{\frac{2\pi}{\rho}\mu}
}
\end{array}
\right),
\end{eqnarray}
where $\hat{G}_0(\beta)
$ is given by (\ref{hG0}).
$\hat{K}(\beta)$ is the general diagonal
boundary $K$-matrix 
associated with $S$-matrix $S(\beta)$
(\ref{def:S}), (See \cite{FK}).
Let us set the state $|\widehat{B}\rangle$
by
\begin{eqnarray}
|\widehat{B}\rangle
=e^{\widehat{B}}|vac\rangle.
\end{eqnarray}
Here we have set
\begin{eqnarray}
\widehat{B}=
\int_0^\infty
\hat{\alpha}_{1,1}(t)a_1(-t)a_1(-t)dt+
\int_0^\infty
\hat{\beta}_1(t)a_1(-t)dt,
\end{eqnarray}
where
\begin{eqnarray}
\hat{\alpha}_{1,1}(t)&=&
-\frac{t}{2}\times
\frac{{\rm sh}\frac{\rho t}{2}}{
{\rm sh}\left(\frac{\rho}{2}+\frac{\pi}{2}
\right)t
}\times
\frac{
{\rm sh}\frac{\pi t}{2}
}{
{\rm sh}\pi t
}.\\
\hat{\beta}_1(t)&=&
-\frac{{\rm sh}\frac{\pi t}{2}}{
{\rm sh}\pi t
}\times
\left(
\frac{{\rm sh}(i\mu-\frac{\rho}{2}
-\frac{\pi}{2})t}{
{\rm sh}\left(\frac{\rho}{2}+\frac{\pi}{2}
\right)t
}+
{\rm sh}\frac{\pi t}{4}
\frac{{\rm ch}\frac{\rho t}{4}}{
{\rm sh}\left(\frac{\rho}{4}+\frac{\pi}{4}
\right)t
}\right).
\end{eqnarray}
We have
\begin{eqnarray}
V_0(\beta)|\widehat{B}\rangle&=&
\hat{G}_0(\beta) V_0(-\beta)|
\widehat{B}\rangle,\\
V_1(\alpha)|
\widehat{B}\rangle&=&
\hat{G}_1(\alpha) V_1(-\alpha)|
\widehat{B}\rangle,
\end{eqnarray}
where
\begin{eqnarray}
\hat{G}_0(\beta)&=&
2^{-\frac{\pi}{2\rho}}
e^{\frac{\gamma}{4}(-1+\frac{\pi}{\rho})}
\times
\frac{\Gamma_2(-2i\alpha+\pi|
\rho,4\pi) \Gamma_2
(-2i\alpha+\rho+3\pi|\rho,4\pi)}
{
\Gamma_2(-2i\alpha|
\rho,4\pi) \Gamma_2
(-2i\alpha+\rho+4\pi|\rho,4\pi)
}\nonumber\\
&&~~~~~~~~~\times
\frac{\Gamma_2(
-i\alpha+i\mu|\rho,2\pi)
\Gamma_2(
-i\alpha-i\mu+\rho+2\pi|\rho,2\pi)
}{\Gamma_2(-i\alpha-i\mu+\rho+\pi|
\rho,2\pi)
\Gamma_2(-i\alpha+i\mu+\pi|
\rho,2\pi)
},
\label{hG0}
\\
\hat{G}_1(\alpha)&=&
-ie^\gamma 2^{1+\frac{\pi}{\rho}}
(\rho e^\gamma)^{1+\frac{\pi}{\rho}-\frac{2i\mu}{\rho}}
\times \alpha 
\times \frac{\Gamma\left(
1+\frac{\pi}{2\rho}+\frac{i(-\mu-\alpha)}
{\rho}
\right)}{
\Gamma\left(-
\frac{\pi}{2\rho}+\frac{i(\mu-\alpha)}
{\rho}
\right)
}.
\end{eqnarray}
As the same arguments as this paper we get
\begin{eqnarray}
\widehat{K}(\beta)_j^j
\widehat{\Psi}_j^*(\beta)|
\widehat{B}\rangle=
\widehat{\Psi}_j^*(-\beta)|\widehat{B}\rangle,
~(j=0,1).
\end{eqnarray}
It seems that
this result is ``homogeneous'' version
of paper \cite{HSWY}.

At last we give some comments 
on $U_q(\widehat{sl_n})$ case.
Let us introduce the operators $\widehat{\Psi}_j^*(\beta),
~(j=0,\cdots,n-1)$ by 
\begin{eqnarray}
\widehat{\Psi}_j^*(\beta)&=&
\int_{C_1^*}d\alpha_1 \cdots
\int_{C_j^*}d\alpha_j :V_0(\beta)
V_1(\alpha_1)\cdots V_j(\alpha_j):\nonumber\\
&\times&
e^{\frac{\pi}{\rho}\beta}
\prod_{k=1}^j
\Gamma\left(
\frac{i(\alpha_{k-1}-\alpha_k)}{\rho}-
\frac{\pi}{n \rho}
\right)
\Gamma\left(
\frac{i(\alpha_{k}-\alpha_{k-1})}{\rho}-
\frac{\pi}{n \rho}
\right).
\end{eqnarray}
The operators
$\widehat{\Psi}_j^*(\beta)$
are slightly different
from
realizations of
$Z_j^*(\beta)$ (\ref{real:Psi*}).
Factor $e^{-\frac{\alpha_j}{\rho}}$
drops in $\widehat{\Psi}_j^*(\beta)$.
It is possible to
construct ``boundary state'' 
$|\widehat{B}\rangle$
for general diagonal boundary $K$-matrix
$
\widehat{K}(\beta)$,
associated with the $S$-matrix $S(\beta)$
(\ref{def:S}),
as the same manner as 
$n=2$ case.
Relating to
general diagonal boundary $K$-matrix,
see Appendix of the paper \cite{FK}.
\begin{eqnarray}
\widehat{K}(\beta)_j^j
\widehat{\Psi}_j^*(\beta)|
\widehat{B}\rangle=
\widehat{\Psi}_j^*(-\beta)|\widehat{B}\rangle,
~(j=0,\cdots,n-1).
\end{eqnarray}
However we cannot derive the commutation
relations (\ref{Com1}) for $n>2$,
under the scheme of the papers
\cite{MT, KY}.
Therefore we select 
the realizations of $Z_j^*(\beta)$
in section 3.
When we consider a limiting case 
$\rho \to \infty$,
two kind of operators $Z_j^*(\beta)$ and 
$\widehat{\Psi}_j^*(\beta)$ become 
free field realizations of
the Zamolodchikov-Faddeev
operators $Z_j^*(\beta)$ given in \cite{K1}.
Therefore, in this limiting case,
we have constructed the boundary state
$|B\rangle$ for general diagonal 
boundary $K$-matrix.
See the papers \cite{K1,FKQ}.

\end{appendix}

\end{document}